%% file: tesePV.tex
\documentclass[brazil,12pt,a4paper,openany]{book}
\usepackage{anysize}
\usepackage[latin1]{inputenc}
\usepackage[english]{babel}
\usepackage[ruled,chapter]{algorithm}
\usepackage{makeidx}
\usepackage{subfig}
\usepackage{amsmath,amssymb,amsthm,mathrsfs}
\usepackage{graphicx}
\usepackage{fancyhdr}
\usepackage{array}
\usepackage{geometry}
\usepackage{simplewick}
\usepackage{latexsym}
\usepackage{float}
\usepackage[all]{xy}
\usepackage{euscript}
\usepackage{enumerate}
\usepackage{dsfont}
\usepackage{slashed}
\usepackage[plainpages=false,pagebackref=true,pdftex]{hyperref}
\usepackage{hyperref}

\usepackage{tabularx,ragged2e,booktabs,caption}
\usepackage{fancyhdr}
\usepackage[nottoc, notlof, notlot]{tocbibind}
\makeindex

\marginsize{25mm}{25mm}{25mm}{25mm}

\begin{document}

%\selectlanguage{english}

\setlength{\abovecaptionskip}{0.0cm}
\setlength{\belowcaptionskip}{0.0cm}
\setlength{\baselineskip}{24pt}

\pagestyle{empty}
\input{capa.tex}                    %
\newpage
\mbox{ }
\pagestyle{headings}

\pagestyle{fancy}
\fancyhf{}      % remove formatações anteriores
\fancyhead[L]{\hfill \thepage}       % poe o nome e n. do cap. e o n. da pagina em cima
\lhead{}
\chead{}
\rhead{\thepage}
\renewcommand{\headrulewidth}{0pt}
\renewcommand{\footrulewidth}{0pt}      % tira o n. da pag. do rodape
\lfoot{}
\cfoot{}
\rfoot{}

\fancypagestyle{plain}
{
	\fancyhf{}
	\lhead{}
	\chead{}
	\rhead{\thepage}
	\lfoot{}
	\cfoot{}
	\rfoot{}
}

\pagenumbering{roman}
\setcounter{page}{2}
\input{assinaturas.tex} 
\input{ficha_catalografica.tex}                    
\input{resumo.tex}                           
\input{abstract.tex}                          
\input{agradecimentos.tex}              
\input{agradecimentos2.tex}                     

\newpage
\phantomsection
\addcontentsline{toc}{chapter}{Summary}
\tableofcontents

\newpage
\phantomsection
\addcontentsline{toc}{chapter}{List of figures}
\listoffigures

\newpage
\phantomsection
\addcontentsline{toc}{chapter}{List of tables}
\listoftables

\newpage
\pagenumbering{arabic}
\setcounter{page}{1} 
\mainmatter
\input{chapter1.tex}
\input{chapter2.tex}
\input{chapter3.tex}
\input{chapter4.tex}
\input{chapter5.tex}
\input{apendicea.tex}
\input{apendiceb.tex}
%****EXTRA*****
%\input{t_x_topicos_para_pesquisar}

\bibliography{tese}{}
\bibliographystyle{plain}

\end{document}

%% file: capa.tex
\frontmatter 

\thispagestyle{empty}

\begin{figure}[h]
	\includegraphics[scale=0.05]{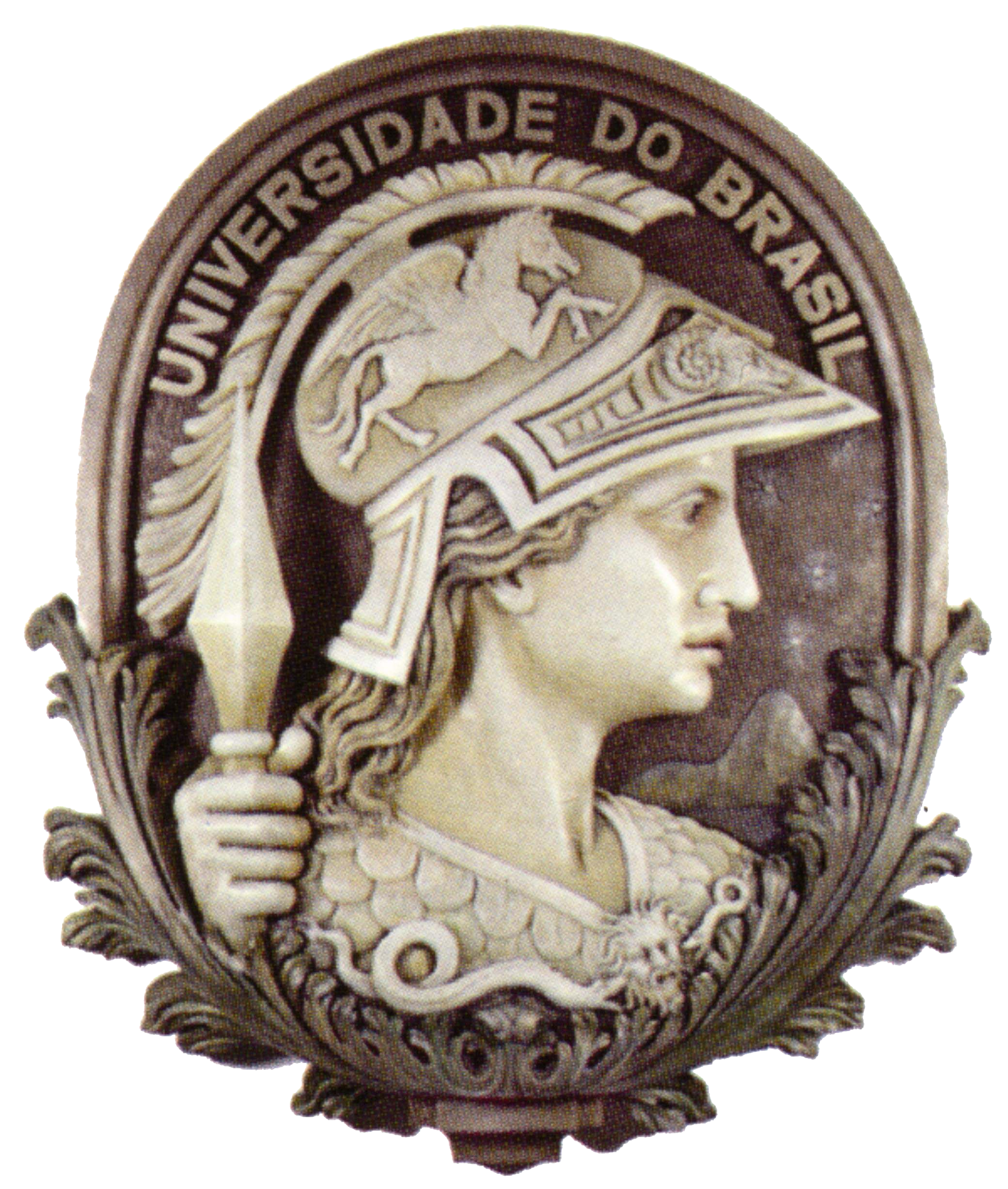}
\end{figure}

\vspace{15pt}

\begin{center}

\textbf{UNIVERSIDADE FEDERAL DO RIO DE JANEIRO}

\textbf{INSTITUTO DE FÍSICA}

\vspace{30pt}

{\Large \bf Electrostatic Force Between Two Colloidal Spheres}

\vspace{25pt}

{\large \bf Daniel Martínez Tibaduiza}

\vspace{35pt}

\begin{flushright}
\parbox{10.3cm}{Tese de Mestrado apresentada ao Programa de Pós-Graduação em Física do Instituto de Física da Universidade Federal do Rio de Janeiro - UFRJ, como parte dos requisitos necessários à obtenção do título de Mestre em Ciências (Física).

\vspace{18pt}

{\large \bf Orientador: Paulo Américo Maia Neto}

\vspace{12pt}

{\large \bf Coorientador: Diney Soares Ether Junior}}
\end{flushright}

\vspace{90pt}

\textbf{Rio de Janeiro}

\textbf{Julho de 2015}

\end{center}

%% file: assinaturas.tex
\newpage

\thispagestyle{empty}

\noindent

%\begin{figure}[h]
%  \begin{center}
%	\includegraphics[scale=0.45]{assinaturas.png}
%  \end{center}
%\end{figure}

\clearpage

%% file: ficha_catalografica.tex
\newpage
\mbox{}\vspace{5cm}

\begin{center}
\begin{tabular}{|c|}
\hline\\
\begin{minipage}{15cm}
\rule{15cm}{0pt} \flushright

\begin{minipage}{12.5cm}
\hspace{2cm} P436 \hspace{0,68cm} Tibaduiza, Daniel Martínez

\hspace{2cm} Electrostatic Force between Two Colloidal Spheres / Daniel Martínez Tibaduiza - Rio de Janeiro: UFRJ/IF, 2015.

\hspace{2cm} xiv, 154f.

\hspace{2cm} Advisor: Paulo Américo Maia Neto

\hspace{2cm} Coadvisor: Diney Soares Ether Junior

\hspace{2cm} Theses (Master) - UFRJ / Instituto de Física / Programa de Pós-graduação em Física, 2015.

\hspace{2cm} Referências Bibliográficas: f. 124-145.

\hspace{2cm}
Mean Words  1. Colloids 2. Two-dielectric spheres 3. Long Range Interactions 3. Double-Layer Electrostatic Force 4. Linearized Poisson-Boltzmann Equation 5. Debye Length.

\bigskip
\end{minipage}
\end{minipage}
\\
\hline
\end{tabular}
\end{center}

%% file: resumo.tex
\newpage

\noindent

\vspace*{20pt}
\begin{center}
{\LARGE\bf Resumo}\\
\vspace{15pt}
{\Large\bf Força Eletrostática Entre Duas Esferas Coloidais}\\
\vspace{6pt}
{\bf Daniel Martínez Tibaduiza}\\
\vspace{12pt}
{\bf Orientador: Paulo Américo Maia Neto}\\
{\bf Coorientador: Diney Soares Ether Junior}\\
\vspace{20pt}
\parbox{14cm}{Resumo da Tese de Mestrado apresentada ao Programa de Pós-Graduação em Física do Instituto de Física da Universidade Federal do Rio de Janeiro - UFRJ, como parte dos requisitos necessários à obtenção do título de Mestre em Ciências (Física).}
\end{center}
\vspace*{35pt}

Nesta dissertação, analisamos a interação de dupla camada entre duas esferas coloidais dielétricas carregadas de raios diferentes imersas em um meio líquido, contendo íons dissociados em água, e em equilíbrio térmico. No limite de potenciais elétricos baixos, a interação é descrita pela equação linearizada de Poisson-Boltzmann (LPBE). Obtivemos uma solução analítica da LPBE para o potencial eletrostático em termos de uma expansão em multipolos supondo densidades superficiais de carga preescritas e uniformes nas esferas. Desenvolvemos um código na plataforma \textbf{Mathematica} que permite calcular a força em função da separação entre as esferas.  Com a finalidade de validar o nosso trabalho, comparamos nossos resultados numéricos com os resultados analíticos válidos na \textit{aproximação de superposição linear} (distâncias muito maiores que o comprimento de Debye) e na \textit{aproximação de força de proximidade} (distâncias e comprimento de Debye muito menores que os raios das esferas). Nosso código será utilizado como parte do modelo teórico para a descrição do experimento de medida da força de Casimir com pinças óticas, atualmente em curso no laboratório de pinças óticas da UFRJ (LPO-COPEA).

\vspace{15pt}

\textbf{Palavras-chave:}  1. Colóides 2. Esferas dielétricas 3. Interações de longo alcance \\4. Força Eletrostática de Dupla Camada 5. Equação de Poisson-Boltzmann Linear  \\6. Comprimento de Debye.

%% file: abstract.tex
\newpage

\noindent

\vspace*{20pt}
\begin{center}
{\LARGE\bf Abstract}\\
\vspace{15pt}
{\Large\bf Electrostatic Force between Two Colloidal Spheres}\\
\vspace{6pt}
{\bf Daniel Martínez Tibaduiza}\\
\vspace{12pt}
{\bf Advisor: Paulo Américo Maia Neto}\\
{\bf Coadvisor: Diney Soares Ether Junior}\\
\vspace{20pt}
\parbox{14cm}{\emph{Abstract} da Tese de Mestrado apresentada ao Programa de Pós-Graduação em Física do Instituto de Física da Universidade Federal do Rio de Janeiro - UFRJ, como parte dos requisitos necessários à obtenção do título de Mestrado em Ciências (Física).} % ATENÇÃO: esse trecho fica em português mesmo!
\end{center}
\vspace*{35pt}
In this dissertation we analyzed the double-layer force interaction between two dielectric charged colloidal spheres of different radii immersed in a solution of ions in water and in thermal equilibrium. In the limit of low electrostatic potential the interaction is governed by the Linearized Poisson-Boltzmann Equation (LPBE). We obtained an analytical solution from the LPBE for the electrostatic interaction via multipole expansion, considering a uniform and fixed surface charge density on the spheres. A code in the \textbf{Mathematica} platform was developed, allowing us to calculate the force between the spheres as a function of their separation. In order to validate the code, we compared our numerical results with the analytical ones in the limits of \textit{Linear Superposition Approximation (LSA)} (valid for sphere separations much greater than the Debye length) and \textit{Proximity Force Approximation (PFA)} (valid for sphere separations and Debye length much smaller than the radii of the spheres). The code will be used as part of the theoretical model of the Casimir force experiment currently in progress at the UFRJ Optical Tweezers Laboratory (LPO-COPEA).

\vspace{15pt}


%% file: agradecimentos.tex
\newpage

\noindent

\vspace*{20pt}

\begin{center}

{\LARGE\bf Acknowledgment}

\end{center}

\vspace*{40pt}

I would initially like to thank the Brazilian people, who through their government; more specifically, the Coordenaçõ de Aperfeiçoamento de Pessoal de Nível Superior (CAPES) foundation, gave me the economic resources and the opportunity to study at the \textit{excellent} Federal University of Rio de Janeiro (UFRJ) and to obtain my \textit{Master degree}. I did my best.
Thanks to Professors Paulo A. Maia and Diney S. Ether for their patience and for their time.
Thanks to the professors and staff of the Physics department at UFRJ and to my friends Ana, Jessica, Leonardo, Duvan, Luis, Diney, Omar, Julio, Jilder, Christopher, Jhonatan and Saulo.\\
For their love which always brings me light and peace thanks to my familiy Mart\'inez, Tibaduiza, Londoño, Willach Galliez, Boileau, Melo, Ospina, Torres, Hernandez, Cristiani Werneck, Pereira, Cabral, Cuyul, Arguellez, Pinillos Valencia, Clavijo, Vargas, and the Suesca Cundinamarca community. 
Thanks to the wonderful city of Rio de Janeiro and its wonderful people that I met.

%% file: agradecimentos2.tex
\newpage

\noindent

\vspace*{20pt}

\begin{center}

\end{center}

\vspace*{40pt}

To my beloved ones\\
											My mothers:	Mercedes, Gilma and Lida.\\
											My fathers: Julio and Fernando.\\
											My brothers: Julio, Gabriel, Bryan and J.M.\\
											My sisters: Alexandra, Lalo, Issis, Carito, Ang\'elica and Tatea.\\
											My wonderful wife Clarita.\\
											My mother and father in law Martina and Carlos.\\
											My nieces and nephews: Louise, Valeria, Mandi, Salom\'e Isabella, Felipe, Simon and Alejo.\\
											
											And particularly to my son, Sebastian, The most beautiful creature that I have ever see.

%% file: chapter1.tex
\begin{chapter}{Introduction}
\label{intro}

\hspace{5 mm} 

When macroscopic charged objects interact, its tendency is to reach a balance such that an object with lesser (higher) quantity of electric charge tends to compensate by adding (losing) charge from other bodies. In the atomic scale atoms this phenomena is characterized by electronegativity. Atoms form molecules and molecules in turn form complex microstructures which form macroscopic matter therefore, which is considered electrically neutral. For instance the tactile sense is the information in our brain that comes from our skin, a certainly electrical neutral object, through electrical impulses telling us that our external surface is in interaction with matter or radiation. Interaction between electrically neutral objects (atoms, molecules and surfaces) during molecular separations is studied in the field of van der Walls forces. 
With the development of quantum mechanics; more specifically the quantization of the electromagnetic field, the description of the van der Waals forces as a consequence of quantum vacuum fluctuations (for a general discussion see \cite{quantumvacuum}) was provided by Hendrik Casimir in 1948. In fact, the explanation of the known phenomena of attraction between neutral parallel plates was reinterpreted by Casimir as zero point energy variations due to the boundary conditions imposed to the field \cite{farina2006}. 
Despite that experiments in the micro and nanoscale that permits the corroboration of this effect has been developed since 1940 \cite{lamoreaux}, only recently the Casimir effect has been measured. In the 40's, the first experiments to understanding the van der Walls forces were realized by Verwey and Overbeek using colloidal systems \cite{Verwey1999}. Colloidal systems have two concurrent phases: one continuous and another, with dimensions smaller than hundreds of micrometers; dispersed in the continuous one. Some actual experiments use the atomic force microscope (AFM) \cite{munday2009,munday2008} to measure the Casimir force in different configurations; however, the distance range is limited by tens of nanometers due to the apparatus limitation \cite{pires}. In 2015, an experiment in the Optical Tweezers Laboratory (LPO-COPEA) at UFRJ was designed to measure the Casimir force between two dielectric colloidal microspheres beyond the Proximity Force Approximation (PFA) regime \cite{probing}. This experiment is the first one of its kind, considering the optical tweezers have never been used to measure Casimir forces. Optical tweezers can measure forces in the femtonewton ($1fN=10^{-15}N$) range. The experimental sample is a colloidal system composed by a solute of charged polystyrene spheres dispersed in a water-salt solution. An optically trapped sphere is approached to the second one, which is attached to the coverslip. In a static frame two forces are expected to act during this process: the Casimir and the double layer electrostatic forces. This dissertation is about the last of these forces. Through the development of a code in the Mathematica platform, we will be able to calculate the exact values for the double layer electrostatic force for different variable values of the fundamental physical characteristics of the experiment, such as Debye screening length, spheres radii and spheres separation. 
Prior to the analysis and results, we briefly discuss some basic introductory ideas. In this light, chapter \ref{cap2} is devoted to an introduction to colloidal systems, the double layer system, and a derivation of the Linear Poisson Boltzmann Equation (LPBE). Important parameters such as the Debye length and bulk ionic concentrations are introduced as well. The LPBE derivation begins to consider the Maxwell equations in the electrostatic regime; which combined with a macro-canonical analysis, leads the Poisson-Boltzmann Equation (PBE). The latter is -then linearized for low electrostatic potentials as in the experimental frame. Additionally, we consider the case of a single dielectric charged colloidal sphere as an example of application of the LPBE. This is relevant since for enough large separations of two charged colloidal spheres, they can be considered without interaction and each potential should be reduced to the single sphere case.
Chapter \ref{cap3} deals with the analysis of the electrostatic interaction between two dielectric colloidal spheres of different radii. Starting from the general solution of the LPBE as a multipole expansion, we use the boundary conditions to find analytical expressions for the coefficients of the expansion that allow us to evaluate the potential numerically. Afterwards, the force between the two spheres as function of it separation is found. In order to compare our numerical results with analytical expressions (for the potential and the force between the spheres) we derive two analytical approaches: the linear superposition approximation (LSA), valid for sphere separations much larger than the Debye length, and the proximity force approximation (PFA), valid for sphere separations and Debye lengths much smaller than the radii of the spheres. 
Chapter \ref{cap4} shows our numerical results of the exact calculation of the force and their comparison with the analytical approach: where the code is validated. We show the behavior of the force between the spheres as function of separation for fixed values of Debye length and different orders of the multipole expansion. The exact numeric results are confronted with the values for the force in the LSA and PFA limits; finding an excellent agreement. 
In the last chapter \ref{cap5}, we present our conclusions and perspectives.

\end{chapter}
\

%% file: chapter2.tex
\begin{chapter}{Foundations}
\newcommand{\scalprod}[2]{\ensuremath{\left\langle{#1}|{#2}\right\rangle}}
\newcommand{\brm}[1]{\boldsymbol{\mathrm{#1}}}
\newcommand{\bsy}[1]{\boldsymbol{{#1}}}
\newcommand{\vect}[1]{\boldsymbol{#1}}
\newcommand{\oper}[1]{\boldsymbol{\mathsf{#1}}}
\newcommand{\vac}{\textsc{vac}}
\newcommand{\sinc}{\ensuremath{\mathrm{sinc}}}
\newcommand{\steve}[1]{{\color{blue} #1}}

\label{cap2}
\hspace{5 mm}

\section{Colloidal systems}
From the colloid definition in $1861$ by the Scottish chemist Thomas Graham as materials that seemed to dissolve but were not able to penetrate a membrane, to our present time, colloidal systems (the current name for colloid) became a strong object of interest, especially in the last $20$ years \cite{Butt}. Colloidal systems are composed by a disperse phase, with dimensions from nanometer to the micrometer scale, and a continuous phase(s)\footnote{By phase we mean a state of matter and we are considering only: solid, liquid and gas.} \cite{Butt}. They are part of our daily life: from technological development, passing from biology and medical science, to clouds in the firmament. The last; for instance, is named aerosol and is a two-phase colloidal system, where liquid and gas are dispersed and the continuous phases respectively. Colloidal systems are classified depending on the phase types involved and different applications and descriptions are available (see table \ref{tab:collo}). 
\begin{table}
	\centering
		\includegraphics[width=16cm]{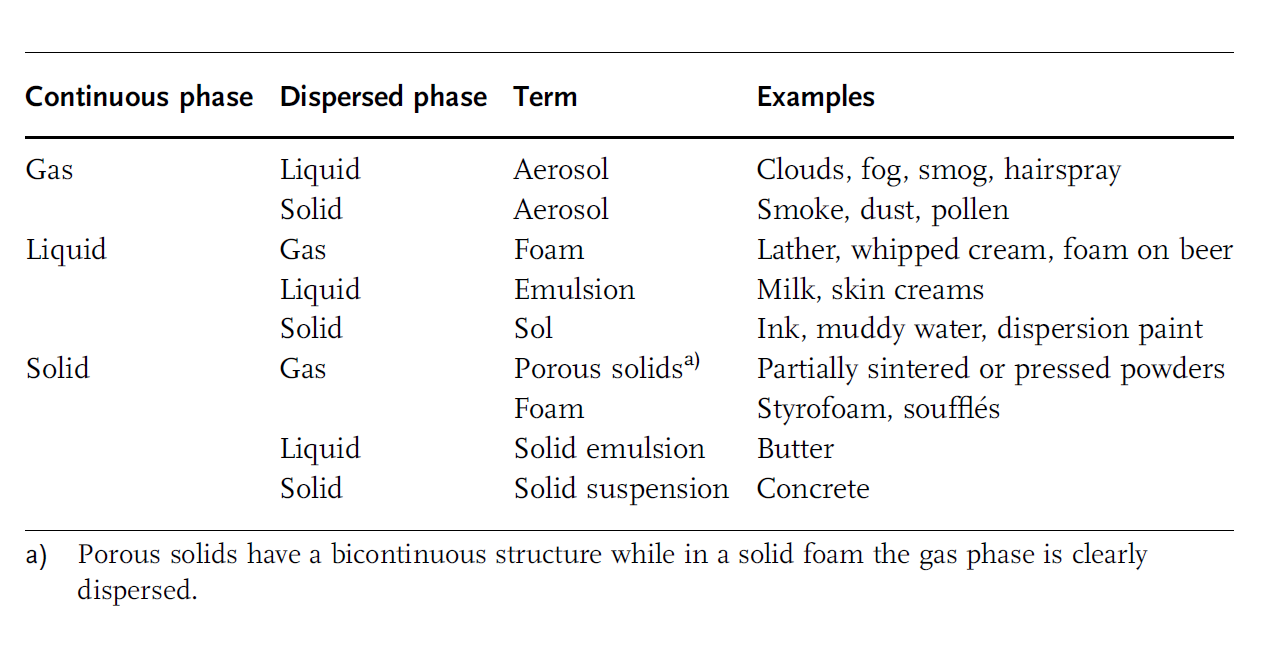} 
	\caption{Types of dispersions. Taken from Ref.\cite{Butt}.}
	\label{tab:collo}
\end{table}
Colloidal chemistry has been the stage of some important works in the field of Van der Waals forces, quantum electrodynamics, and double-layer interactions. For instance, Overbeek and Verwey developed the initial theory about stability of colloidal suspensions (quartz dust particles) for the Philips Company in the 40's \cite{verwey99}, relating Van der Waals and electrostatic double-layer forces, and obtaining an interaction between particles of the order of $~1/r^{6}$. However, this theory was not in agreement with the experimental data where the order of the interaction decays as $~1/r^{7}$. It was only until the development by Casimir and Polder in the context of perturbative quantum electrodynamics (QED), when retardation effects of the electromagnetic interaction were taken into account; thus, the theory predictions and experimental results agreed. In fact, the seminal work of Casimir using the frame of Quantum Electrodynamics (variation of the zero point energy) to describe this type of interaction effects between two perfectly conductor parallel plates, leads to substantial consequences as a quantum vacuum fluctuations experimental testing, and a way to validate quantum electrodynamics via the Casimir Effect \cite{bordag1999}.
The colloidal system of our interest is a Sol. It is a lyophobic colloid \footnote{When the components of the disperse phase do not have affinity with the molecules of the dispersion medium \cite{verwey99}, \emph{i.e.} the solute constituents do not swell (hydration for instance).} composed by a solid disperse phase of two naturally charged \cite{verwey99, ohara1984} polystyrene spheres of radii $ 7.2\mu m$ and $ 1.5\mu m$, dispersed in a water-monovalent salt solution liquid. Due to the high water dielectric constant, the salt is -dissociated in its ions and counter-ions, which in the solution are initially homogeneously distributed.

\section{The Double Layer}

\begin{figure}[H]
\centering
\includegraphics[width=10cm]{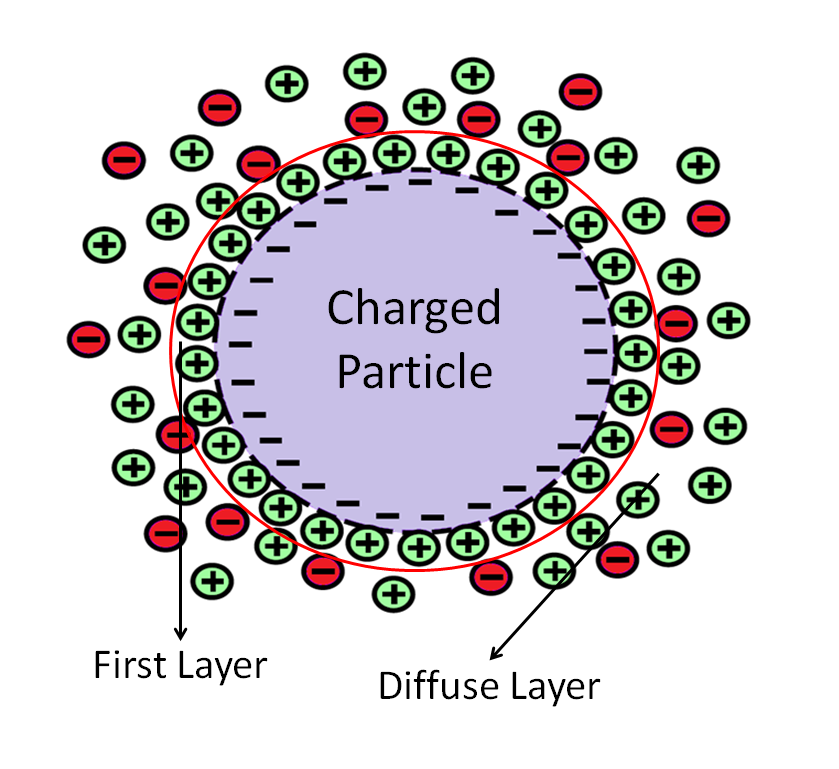} 
\caption{Double-Layer Scheme. The sizes are not in scale.}
\label{fig:DL}
\end{figure}

\textit{`From several phenomena observed in colloidal systems it has been inferred that the dispersed particles carry an electric charge'} \cite{verwey99}. In fact, from old experiments in electrostatic, we can recall how plastic is easily charged using fleece cloth. Polystyrene is naturally charged also, and even when the residual charge is removed via immersion in ethyl alcohol and air-dried, a residual charge of the order of $25 nC/m^{2}$ remains on the surface \cite{ohara1984}. According to Ref.\cite{verwey99}: \textit{`Though in reality it is a charge consisting of point charges, it is customary to consider it, as a first approximation, as a homogeneous surface charge spread over the surface of the particles'}. As a result, when the solution ions; which are initially homogeneously distributed, interact with this residual surface charge density, they redistribute themselves creating a diffuse layer around the spheres. The sphere surface charge density with the primary adsorbed ions forms the first layer, and together with the second diffuse one, constitute a system known as the double layer (see Figure.\ref{fig:DL}), which has been studied from the beginning of the twentieth century until today \cite{Butt}.

The double layer force together with the Van der Waals-Casimir interaction between the spheres will support the colloidal system stability. In fact, it is remarkable that this simple picture, \emph{i.e.} equilibrium given by only two forces, proposed by Derjaguin and Landau, and Verwey and Overbeek (DLVO theory) \cite{tadros2007}, is sufficient to describe the colloidal stability. In a general sense, the system stability depends on: (i) ion concentration, since a huge ion concentration screens the double-layer repulsive interaction, favoring the attractive Van der Waals interaction and giving rise to an agglomeration state. (ii) The system temperature, as a second parameter that permits to restrain the energy of the electrostatic interactions between the ions. Thirdly,  (iii) the solute particle sizes and its component types. (iv) Lastly, the solvent properties. Additionally, it is important to note that the stability state also depends on the observation time scale. 

In the last century, different models were created to describe some of the double-layer characteristics. For instance, considering planar surfaces in the simplest proposed model (Helmholtz Layer) \cite{Butt} the ions layer was considered directly binded to the solute surface; neutralizing the first layer. Models that are more sophisticated includes thermal motion as was first proposed by Gouy and Chapman (see Fig.\ref{fig:BD}) for the planar surfaces. In this thesis, we consider the Debye-Huckel model \cite{DebyeH1923} that takes into account the thermal motion and the associated ionic distribution for solute elements with spherical geometry.

\begin{figure}[H]
\centering
\includegraphics[width=10cm]{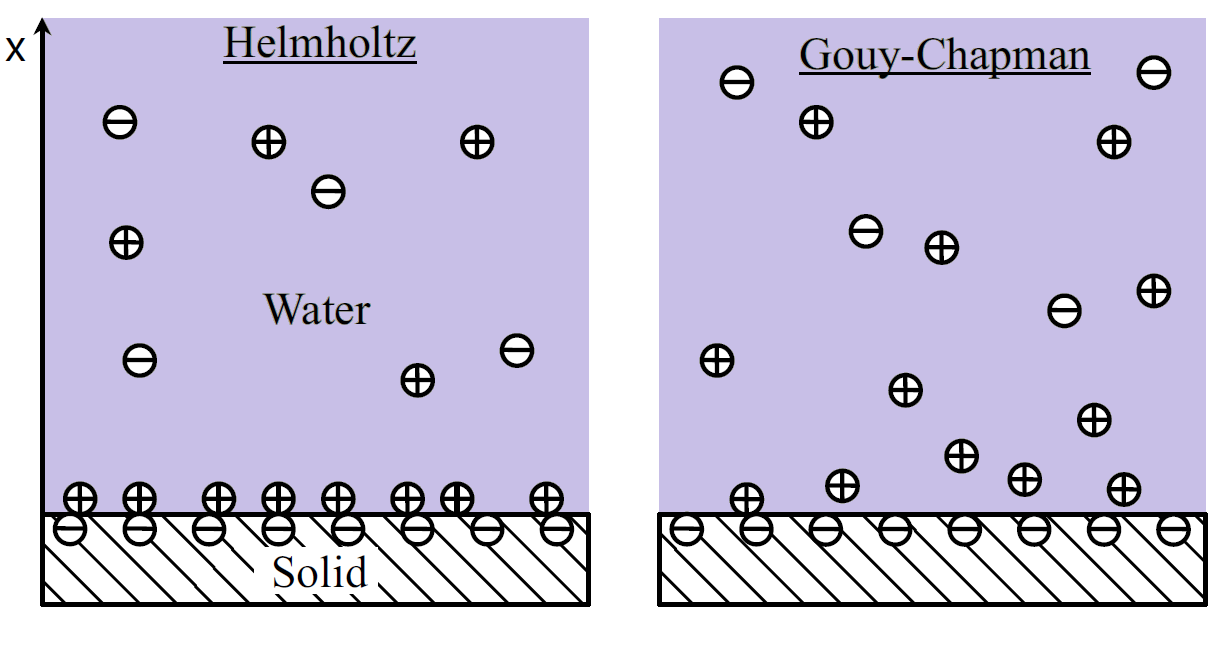} 
\caption{Helmholtz and Gouy-Chapman model of the electric double layer for planar surfaces. Taken from \cite{Butt}.}
\label{fig:BD}
\end{figure}

%%%%%%%%%%%%%%%%%%%%%%%%%%%%%%%%%%%%%%%%%%%%%%%%%%%%%%%%%%%%%%%%%%%%%%%%%%%%%%%%%%%%%%%%%%%%%%%%%%%%%%%%%%%%%%%%%%%%%

\section{The Linearized Poisson-Boltzmann Equation (LPBE)}
\subsection{Derivation of the LPEB}
In a general sense, the solute elements (polystyrene spheres in our case) are treated as dielectric macroscopic objects. Therefore, we consider the Maxwell equations (\textit{S.I} system) in the electrostatic regime \cite{Jackson98}:
\begin{equation}
\nabla\cdot\textbf{D}=\rho
\label{eq:Max1}
\end{equation}
\begin{equation}
\nabla\times\textbf{E} = 0
\label{eq:Max2}
\end{equation}
where $\textbf{D}$ is the displacement vector, $\rho$ is the free charge in the solution and $\textbf{E}$, the electric field. For a linear and isotropic solvent media $\textbf{D}=\epsilon \textbf{E}$, where $\epsilon$ is the electric permittivity of the solvent \cite{Jackson98}. From Eq.(\ref{eq:Max2}) the electric field can be written as the gradient of a scalar field: 
\begin{equation}
\textbf{E}(x,y,z) =-\nabla\psi(x,y,z)
\label{eq:epsi}
\end{equation}
with $\psi$ the electric potential. Using $\textbf{D}=\epsilon \textbf{E}$ and Eq.(\ref{eq:epsi}) in Eq.(\ref{eq:Max1}) we derive the Poisson equation
\begin{equation}
\nabla^{2}\psi=-\frac{\rho}{\epsilon} \,
\label{eq:Poisson}
\end{equation}
which governs the electrostatic potential outside the macroscopic dielectric objects (in the solution). Since there is not charge inside the macroscopic dielectric objects ($\rho=0$), the Laplace equation governs the electrostatic potential 
\begin{equation}
\nabla^{2} \psi=0.
\label{eq:Laplace}
\end{equation}
The colloidal system is considered stable and adiabatically isolated, \emph{i.e.} the total energy and solvent particle number fixed. Following Debye and Huckel \cite{richet2012}, the following additional assumptions are adopted: ($i$) the ions are treated as point charges that do not accumulate and they generate a symmetrical Coulomb field; ($ii$) the mutual electrostatic energy of two ions in their closest distance of approach is smaller compared to their average thermal energy and ($iii$) the presence of the ions has no effect on the dielectric constant of the solvent. For a $z:z$ dissociation\footnote{A $z:z$ electrolyte is a substance which separates into cations and anions of $z$ and $-z$ valence, respectively.} the chemical potential of a ($\pm$) type of ion \cite{israelachvili11} may be written as
\begin{equation}
\mu=\pm\textit{e}z\psi+k_{B}T\log{n^{\pm}} \,\,
\label{eq:chemicalpot}
\end{equation}
where $n^{\pm}$ is the number density of ($\pm$) ions of valency $z$ at any point $(x,y,z)$ between two solute surfaces. Since in equilibrium the chemical potential is required to be the same \cite{israelachvili11}, from Eq.(\ref{eq:chemicalpot}) is obtained the local ionic distribution (or local ion density\cite{Butt})
\begin{equation}
n^{\pm}=n_{0}e^{\mp\frac{\textit{e}z\psi}{k_{B}T}}\,\,
\label{eq:iondensity}
\end{equation}
where $n_{0}$ is known as the ionic bulk concentration and is the ion density in regions where its distribution is not affected by the electrostatic potential $\psi$. From the two-ion type contribution, the local charge density is:
\begin{equation}
\rho=z\textit{e}n^{+}+(-z\textit{e})n^{-}=\textit{e}z\textit{n}_{0}\left(e^{-\frac{\textit{e}z\psi(x,y,z)}{k_{B}T}}-e^{\frac{\textit{e}z\psi(x,y,z)}{k_{B}T}}\right)\,.
\label{eq:IonDis}
\end{equation}
When Eq.(\ref{eq:IonDis}) is plugged into the Poisson equation (\ref{eq:Poisson}) we obtain the Poisson-Boltzmann equation, which describes the behavior of the electrostatic potential in the solvent medium, \emph{i.e.} outside the colloidal particles:
\begin{equation}
\nabla^{2} \psi=\frac{\textit{e}z\textit{n}_{0}}{\epsilon}\left(e^{\frac{\textit{e}z\psi}{k_{B}T}}-e^{-\frac{\textit{e}z\psi}{k_{B}T}}\right)
\label{eq:PoissonBoltzmann}
\end{equation}
this is a nonlinear second order partial differential equation where the right side is proportional to the hyperbolic sine function
\begin{equation}
\nabla^{2} \psi=\frac{2\textit{e}z\textit{n}_{0}}{\epsilon}\sinh\left(\frac{\textit{e}z\psi}{k_{B}T}\right)\,.
\label{eq:PBH}
\end{equation}
The solution of this equation depends on the geometry of the system and since the Laplace equation is not separable for all coordinates, there is no analytical solution for all geometries. Examples of analytical solutions for a planar surface and a few additional geometries can be found in Ref.\cite{Butt}.
We are interested in the regime of low electrostatic energies, $\textit{e} \left|\psi\right|\ll k_{B}T$, \emph{i.e.} the case when the electrostatic potential is much smaller than the thermal potential $\frac{k_{B}T}{\textit{e}}$ which at room temperature is approximately 25$mV$ \footnote{However, Ref.\cite{Chan94} indicates that under certain circumstances, the LPBE gives accurate results for potentials up to about 40$mV$}. With this in mind, we expand the hyperbolic function as a Taylor series around zero:
\begin{equation}
\sinh\left(\frac{\textit{e}z\psi}{k_{B}T}\right)=\frac{\textit{e}z\psi}{k_{B}T}+{\cal O}\left(\frac{\textit{e}z\psi}{k_{B}T}\right)^{2} \,.
\label{eq:TS}
\end{equation}
By neglecting terms of higher order than 2, we obtain the linearized Poisson-Boltzmann equation (LPBE)
\begin{eqnarray}
\nabla^{2}\psi&=&\frac{2\textit{e}z\textit{n}_{0}}{\epsilon}\cdot \frac{\textit{e}z\psi}{k_{B}T}\,\,\nonumber\\
\nabla^{2}\psi&=&\kappa^{2}\psi \,\,
\label{eq:LPBE}
\end{eqnarray}
where $\kappa=\sqrt{\frac{2(\textit{e}z)^{2}\textit{n}_{0}}{\epsilon k_{B}T}}$ is the inverse of the Debye length $\lambda_{D}=\frac{1}{\kappa}$, which is the length scale characterizing the thickness of the charge distribution (diffuse layer) as we shall see in the next section. Therefore, it is important to note that the local charge density is linearly proportional to the electrostatic potential. From Eqs.(\ref{eq:LPBE} and \ref{eq:Poisson})
\begin{equation}
\rho=-\epsilon\kappa^{2}\psi
\label{eq:localdenpot}
\end{equation}
%%%%%%%%%%%%%%%%%%%%%%%%%%%%%%%%%%%%%%%%%%%%%%%%%%%%%%%%%%%%%%%%%%%%%%%%%%%%%%%%%%%%%%%%%%%%%%%%%%%%%%%%%%%%

\subsection{An example: a single dielectric colloidal sphere}

\label{sec:example}
\begin{figure}[H]
\centering
\includegraphics[width=12cm]{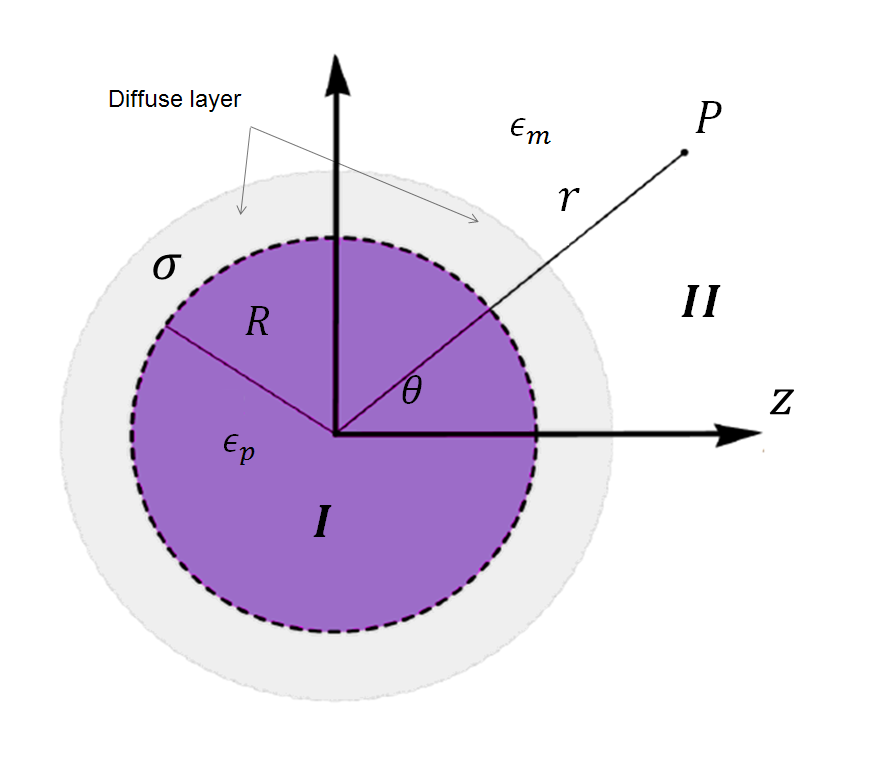} 
\caption{Colloidal sphere scheme}
\label{fig:OS}
\end{figure}
Consider a homogeneous and isotropic dielectric sphere of radius $R$ with relative electrical permittivity $\epsilon_{p}$ and uniform surface charge density $\sigma$, in a $z:z$ solution in thermal equilibrium and relative electrical permittivity $\epsilon_{m}$(see Fig.\ref{fig:OS}). For low energies, the electrostatic potential satisfies the LPBE (\ref{eq:LPBE}) outside the sphere and the Laplace equation (\ref{eq:Laplace}) inside it: 
\par
\begin{eqnarray}
\nabla^{2}\psi(P)&=&\kappa^{2}\psi(P)\,\,\,\,\mbox{(if $P \in I\!I$)}\nonumber\\
&=&0\,\,\,\,\mbox{(if $P \in I$)}
\label{eq:PBlO}
\end{eqnarray}
\par
where
\begin{equation}
\kappa=\sqrt{\frac{2n_{0}(ez)^{2}}{\epsilon_{m}\epsilon_{0} k_{B}T}} \,.
\label{eq:invDebye}
\end{equation}
By the system conditions, we may associate spherical symmetry in the electrostatic potential: $\psi=\psi(r)$, where $r$ is the distance from the center of the sphere to the point of evaluation. Writing the LPBE in spherical coordinates, we obtain
\begin{equation}
\frac{1}{r^{2}}\frac{\partial}{\partial r}\left(r^{2}\frac{\partial \psi}{\partial r}\right)=\kappa^{2}\psi \,.
\label{eq:LapSph}
\end{equation}
The general solution of (\ref{eq:LapSph}) is
\begin{eqnarray}
\psi=A\frac{e^{-\kappa r}}{r}+B\frac{e^{\kappa r}}{r}\,\,\,\mbox{(if $P \in I\!I$)}
\label{eq:SolLapSph}
\end{eqnarray}
in order to have a finite potential at $r\rightarrow \infty$ in Eq.\ref{eq:SolLapSph} we should take $B=0$. Inside the sphere we have
\begin{eqnarray}
\frac{\partial}{\partial r}\left(r^{2}\frac{\partial \psi}{\partial r}\right)&=&0\,\,\,\mbox{(if $P \in I$)}\nonumber\\
\frac{d \psi}{d r}&=&\frac{C}{r^{2}}\,\,\nonumber\\
\psi&=&-\frac{C}{r}+D \,\,.
\label{eq:LapSphIn}
\end{eqnarray}
Since $\psi(r)$ must be finite at the origin, we take $C=0$. The task to find the coefficients $A$ and $D$ is accomplished by using the boundary conditions
\begin{eqnarray}
\left.\psi^{I}(r)\right|_{r=R^{-}}=\left.\psi^{I\!I}(r)\right|_{r=R^{+}}&&\\
\left[\left.\epsilon_{p}\nabla\psi^{I}(r)\right|_{r=R^{-}}-\left.\epsilon_{m}\nabla\psi^{I\!I}(r)\right|_{r=R^{+}}\right]\cdot \hat{\vect{n}}&=&\frac{\sigma}{\epsilon_{0}}\nonumber\\
&&
\label{eq:bcond}
\end{eqnarray}
where $\hat{\textbf{n}}$ is the unit vector normal to the sphere surface. From the boundary conditions we obtain the system of equations
\begin{eqnarray}
D&=&A\frac{e^{-\kappa R}}{R}\,\,\nonumber\\
\epsilon_{m}A \frac{e^{-\kappa R}}{R^{2}}(\kappa R+1)&=&\frac{\sigma}{\epsilon_{0}}
\label{eq:Line}
\end{eqnarray}
and then
\begin{eqnarray}
A&=&\frac{\sigma}{\epsilon_{m}\epsilon_{0}}e^{\kappa R}\left(\frac{R^{2}}{1+\kappa R}\right)\,\,\nonumber\\
D&=&\frac{\sigma}{\epsilon_{m}\epsilon_{0}}\left(\frac{R}{1+\kappa R}\right)=:\psi_{0} \,.
\label{eq:LineSol}
\end{eqnarray}
The potential is finally given by:
\begin{eqnarray}
\psi(r)&=&\psi_{0}R e^{\kappa R}\left(\frac{e^{-\kappa r}}{r}\right)\,\,\,\mbox{(if $P \in I\!I$)}\nonumber\\
\psi(r)&=&\psi_{0}\,\,\,\mbox{(if $P \in I$)} \,.
\label{eq:allpot}
\end{eqnarray}
From Eq.\ref{eq:localdenpot}, the local charge density distribution is proportional to the potential in the solvent
\begin{eqnarray}
\rho(r)&=&-\epsilon_{0}\kappa^{2}\psi(r)\,\,\nonumber\\
&=&\rho_{0}R e^{\kappa R}\left(\frac{e^{-\kappa r}}{r}\right)
\label{eq:localden}
\end{eqnarray}
where
\begin{equation}
\rho_{0}=:-\frac{\kappa^{2}\sigma}{\epsilon_{m}}\left(\frac{R}{1+\kappa R}\right)
\label{eq:ro0}
\end{equation}

\begin{figure}[H]
\centering
\includegraphics[width=17cm]{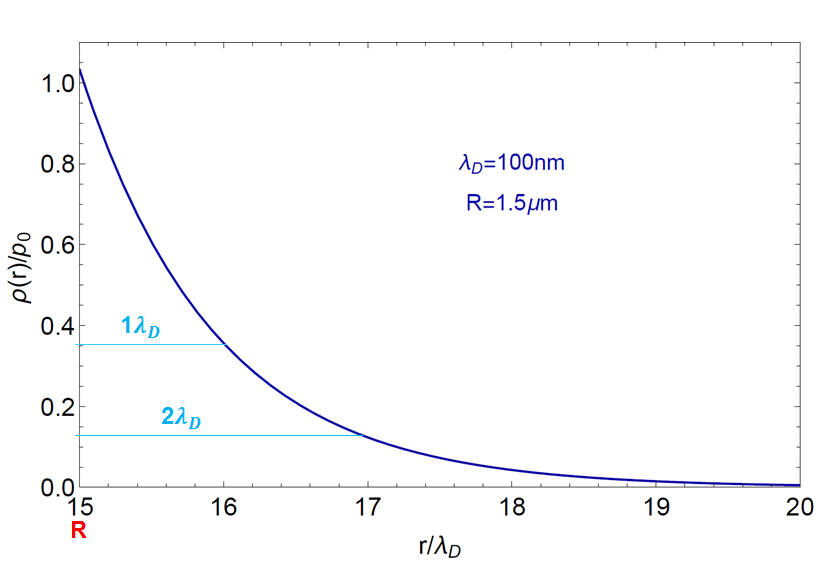} 
\caption{Ratio of the ionic charge density for a single colloidal sphere of radius $R=1.5\mu m$ and Debye \textit{screening} lenght $\lambda_{D}=100nm$.} \,.
\label{fig:OS}
\end{figure}

In Fig.\ref{fig:OS} the ion distribution is plotted around a dielectric sphere of radius $R$ in a salt-water solution for a typical Debye length value ($\lambda_{D}=100nm$) \cite{israelachvili11}. The radius of the sphere was given by R=1.5um, which is a typical value used in optical tweezers experiments. The ion concentration decays rapidly away from the edge. This decay is characterized by the Debye length $\lambda_{D}$; thus also the attenuation of the potential. For $r>\lambda_{D}$ the potential and ion concentration decays substantially (more than $60\%$) whereas for $r>2\lambda_{D}$ it decays almost $90\%$. 

\end{chapter}

%% file: chapter3.tex
\begin{chapter}{Double Layer Interaction Between Two Colloidal Spheres}
\newcommand{\scalprod}[2]{\ensuremath{\left\langle{#1}|{#2}\right\rangle}}
\newcommand{\brm}[1]{\boldsymbol{\mathrm{#1}}}
\newcommand{\bsy}[1]{\boldsymbol{{#1}}}
\newcommand{\vect}[1]{\boldsymbol{#1}}
\newcommand{\oper}[1]{\boldsymbol{\mathsf{#1}}}
\newcommand{\vac}{\textsc{vac}}
\newcommand{\sinc}{\ensuremath{\mathrm{sinc}}}
\newcommand{\steve}[1]{{\color{blue} #1}}

\label{cap3}
\hspace{5 mm} 

In this chapter, the interaction between two charged colloidal dielectric spheres of different radii is analyzed.  The general analytical expressions for the electrostatic potential and the force between them are obtained, together with the corresponding results for two important limits.

\section{Electrostatic Potential Between Two Dielectric Spheres}

\begin{figure}
\centering
\includegraphics[width=15cm]{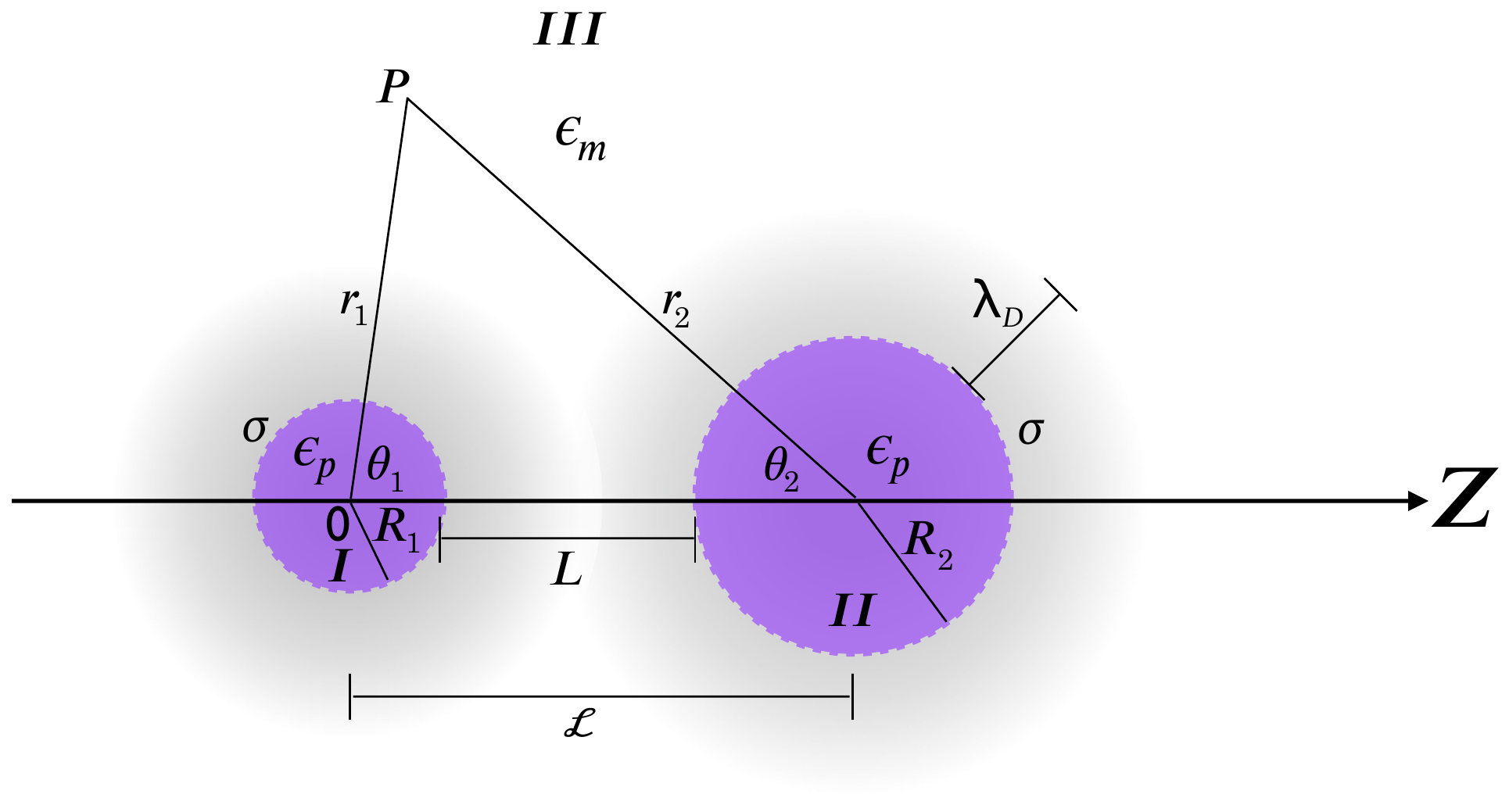} 
\caption{Two dielectric colloidal spheres system with its physical attributes.}
\label{fig:two-spheres}
\end{figure}
From the preceding chapter, we saw that the LPBE governs the electrostatic potential outside dielectric matter concentrations with linear response, immerse in a monovalent solution with the free charge following a Boltzmann distribution. 
The system now is two charged spheres of different radii but with the same known relative permittivity, immerse in a dielectric $z:z$ electrolytes (for instance, NaCl is 1:1 electrolyte) solution in thermal equilibrium (at room temperature $T=298K$). The smallest distance between the spheres is $L$, \emph{i.e.} from its edges. The spheres surface charge is uniform, prescribed and fixed for all spheres separations. Our task is to find the force in one sphere due to the solution electrolytes and the electrical surface charge density of the other sphere. By this the potential in all the space is the first objective.
In the example \ref{sec:example}, by the spherical geometry of the system, the potential was  only a function of the radius. However, there is no more spherical symmetry for the two-spheres system (Figure \ref{fig:two-spheres}), but still an azimuthal one, \emph{i.e.} the electrostatic potential is $\phi$ independent. Again, the potential satisfies the equations
\par
\begin{eqnarray}
\nabla^{2}\psi(P)&=&\kappa^{2}\psi(P)\,\,\,\,\mbox{(if $P \in I\!I\!I$)}\nonumber\\
&=&0\,\,\,\,\mbox{(if $P \in I$ and $I\!I$)}
\label{eq:PBlO}
\end{eqnarray}
\par
where
\begin{equation}
\lambda_{D}=\frac{1}{\kappa}=\sqrt{\frac{\epsilon_{m}\epsilon_{0} k_{B}T}{2n_{0}(ez)^{2}}} \,\,
\label{eq:invDebye}
\end{equation}
is the Debye length, which is a measurement of the diffuse layer spatial decay \cite{Butt}.

%%%%%%%%%%%%%%%%%%%%%%%%%%%%%%%%%%%%%%%%%%%%%%%%%%%%%%%%%%%%%%%%%%%%%%%%%%%%%%%%%%%%%%%%%%%%%%%%%%%%%%%%%

\subsection{General solution inside and outside the spheres}
\subsubsection{\textbf{Outside the spheres:}}
\par The Helmholtz equation
\begin{equation}
\nabla^{2}\psi+\kappa_{H}^{2}\psi=0 \,\,
\label{eq:Helmholtzeq}
\end{equation} 
have the well-known solution in multipole expansion for spherical coordinates \cite{arfken2005}
\begin{equation}
\psi^{H}=\sum_{l,m}{a_{lm}\psi^{H}_{lm}} \,\,
\label{eq:HelmholtzSol}
\end{equation}
with
\begin{eqnarray}
\psi_{lm}^{H}(r,\theta,\phi)=\left\{
\begin{array}{ll}j_{l}(\kappa_{H}r)\\
y_{l}(\kappa_{H}r)\\
\end{array}\right\}
\left\{
\begin{array}{ll}P^{m}_{l}(\cos \theta)\\
Q^{m}_{l}(\cos \theta)\\
\end{array}\right\}
\left\{
\begin{array}{ll}\cos m\phi\\
\sin m\phi\\
\end{array}\right\}^{b}\,\,
\label{eq:HelmholtzSol}
\end{eqnarray}
where $j_{l}$ and $y_{l}$ are the spherical Bessel functions of the first and second kind respectively and $P^{m}_{l}$ and $Q^{m}_{l}$ are the $l$ order regular and irregular associated Legendre polynomial. When $\kappa_{H}$ is a imaginary pure number with imaginary part $\kappa$, the Helmholtz equation becomes the LPBE
\begin{eqnarray}
\nabla^{2}\psi+\kappa_{H}^{2}\psi&=&0 \,\,\nonumber\\
\rightarrow \nabla^{2}\psi+(i\kappa)^{2}\psi&=&0 \,\,\nonumber\\
\rightarrow \nabla^{2}\psi-\kappa^{2}\psi&=&0 \,\,\,.
\label{eq:HeltoLPBE}
\end{eqnarray} 
As a result, the LPBE solution is a modification of the Helmholtz one Eq.(\ref{eq:HelmholtzSol}). Therefore, using the multipole expansion in spherical coordinates the solution of the LPBE is
\begin{equation}
\psi=\sum_{l,m}{a_{lm}\psi_{lm}} \,\,
\label{eq:LPBEGSol}
\end{equation}
with
\begin{eqnarray}
\psi_{lm}(r,\theta,\phi)=\left\{
\begin{array}{ll}i_{l}(\kappa_{H}r)\\
k_{l}(\kappa_{H}r)\\
\end{array}\right\}
\left\{
\begin{array}{ll}P^{m}_{l}(\cos \theta)\\
Q^{m}_{l}(\cos \theta)\\
\end{array}\right\}
\left\{
\begin{array}{ll}\cos m\phi\\
\sin m\phi\\
\end{array}\right\}^{b}\,\,
\label{eq:LPEBSol}
\end{eqnarray}
where $i_{l}$ and $k_{l}$ are the modified spherical Bessel functions of the first and second kind respectively and they are related to the modified Bessel function of half integer order: $I_{n+\frac{1}{2}}$ and $K_{n+\frac{1}{2}}$, by $i_{n}(x)=(\pi/2x)^{1/2}I_{n+\frac{1}{2}}(x)$ and $k_{n}(x)=(\pi/2x)^{1/2}K_{n+\frac{1}{2}}(x)$ \cite{abramowitz64}. By the azimuthal symmetry condition, we may put $m=0$, and since the irregular Legendre solutions have a divergent behavior \cite{arfken2005}, they are not useful to describe our problem, so we may write $Q^{0}_{l}=0$. By the condition that the potential be zero in the infinite, we make $i_{l}=0$. Hence the solution in the medium for the two spheres can be written as a expansion from its centers \cite{Chan94} as:
\begin{equation}
\psi^{I\!I\!I}(P)=\sum_{n=0}^{\infty}\left[a_{n}k_{n}(\kappa r_{1})P_{n}(\cos \theta_1)+b_{n}k_{n}(\kappa r_{2})P_{n}(\cos \theta_2)\right]\,\,.
\label{eq:psiout1}
\end{equation}
This solution no need depend of the four variables ($r_{1},r_{2},\theta_{1},\theta_{2}$) that belongs to a triangle and can be reduced to only three (see Fig.\ref{fig:two-spheres}). Therefore, in order to compare them with the potential inside the spheres via boundary conditions, we can use the addition theorem for Bessel functions \cite{gray1895} to write the Eq.\ref{eq:psiout1} as a function only of one radius, one angle, and the sphere center separation
\begin{eqnarray}
\psi^{I\!I\!I}(P)&=&\sum_{n=0}^{\infty}\left[a_{n}k_{n}(\kappa r_{1})P_{n}(\cos \theta_1)+b_{n}\sum_{m=0}^{\infty}(2m+1)B_{nm}(\kappa \mathcal{L})i_{m}(\kappa r_{1})P_{m}(\cos\theta_1)\right]\nonumber\\
&=&\sum_{n=0}^{\infty}\left[a_{n}\sum_{m=0}^{\infty}(2m+1)B_{nm}(\kappa \mathcal{L})i_{m}(\kappa r_{2})P_{m}(\cos\theta_2)+b_{n}k_{n}(\kappa r_{2})P_{n}(\cos \theta_2)\right]\nonumber\\
\label{eq:psiout}
\end{eqnarray}
where
\begin{equation}
B_{nm}(\kappa \mathcal{L})=\sum_{\nu=0}^{\infty}A_{nm}^{\nu}k_{n+m-2\nu}(\kappa \mathcal{L})
\label{eq:Bcoef}
\end{equation} 
with
\begin{eqnarray}
A_{nm}^{\nu}&=&\frac{\Gamma(n-\nu+\frac{1}{2})\Gamma(m-\nu+\frac{1}{2})\Gamma(\nu+\frac{1}{2})}{\pi\Gamma(n+m-\nu+\frac{3}{2})}\times\nonumber\\
&&\frac{(n+m-\nu)!}{(n-\nu)!(m-\nu)!\nu!}\left(n+m-2\nu+\frac{1}{2}\right)\,.
\label{eq:Agamma}
\end{eqnarray} 
\par
In Eq.(\ref{eq:psiout}) $a_{n}$ and $b_{n}$ are unknown coefficients which will be found by applying the appropriate boundary conditions.
\par
\subsubsection{\textbf{Inside the spheres:}}
\par
Following Ref.\cite{griffiths99}, the general solutions of the Laplace equation for azimuthal symmetry in regions $I$ and $\!I\!I$  are respectively given by
\begin{eqnarray}
\psi^{I}(r_{1},\theta_{1})&=&\sum_{n=0}^{\infty}\left(c_{n}r_{1}^{n}+\frac{B_{n}}{r_{1}^{n+1}}\right)P_{n}(\cos \theta_{1})\\
\psi^{I\!I}(r_{2},\theta_{2})&=&\sum_{n=0}^{\infty}\left(d_{n}r_{2}^{n}+\frac{C_{n}}{r_{2}^{n+1}}\right)P_{n}(\cos \theta_{2})\,.
\end{eqnarray} 
Since there is no charge in the sphere centers, $\psi^{I}$ and $\psi^{II}$ must be finite at $r_{1}=0$ and $r_{2}=0$, therefore $B_{n}=C_{n}=0,\,\,\forall n \in\mathbb{N}$. As a result, the corresponding general solutions in multipole expansion in regions $I$ and $\!I\!I$ will be \cite{chan93}:
\begin{eqnarray}
\psi^{I}(r_{1},\theta_{1})&=&\sum_{n=0}^{\infty}c_{n}r_{1}^{n}P_{n}(\cos \theta_{1})\\
\psi^{I\!I}(r_{2},\theta_{2})&=&\sum_{n=0}^{\infty}d_{n}r_{2}^{n}P_{n}(\cos \theta_{2})
\label{eq:psiin}
\end{eqnarray}
where $c_{n}$ and $d_{n}$ are the corresponding unknown coefficients.
%%%%%%%%%%%%%%%%%%%%%%%%%%%%%%%%%%%%%%%%%%%%%%%%%%%%%%%%%%%%%%%%%%%%%%%%%%%%%%%%%%%%%%

\subsection{Boundary Conditions}
\label{sec:boundconddd}
Assuming that the surface charge densities on the spheres $\sigma$ remain fixed for any particle separations, the electrostatic potential satisfies the following boundary conditions \cite{Jackson98}: 
\begin{eqnarray}
\left.\psi^{I}(r_{1},\theta_{1})\right|_{r_1=R_1^{-}}=\left.\psi^{I\!I\!I}(r_{1},\theta_{1})\right|_{r_1=R_1^{+}}&&\\
\left[\left.\epsilon_{p}\nabla\psi^{I}(r_{1},\theta_{1})\right|_{r_1=R_1^{-}}-\left.\epsilon_{m}\nabla\psi^{I\!I\!I}(r_{1},\theta_{1})\right|_{r_1=R_1^{+}}\right]\cdot \hat{\vect{n}}&=&\frac{\sigma}{\epsilon_{0}}\nonumber\\
&&
\label{eq:bcond}
\end{eqnarray}
where $\epsilon_{m}$ and $\epsilon_{p}$ are, respectively, the relative electrical permittivity for the solvent and the dielectric spheres, and $\boldsymbol{\hat{n}}$ is the normal unit vector pointing outward spheres. For sphere 1, the first boundary condition gives
\par 
\small
\begin{equation}
\sum_{n=0}^{\infty}\left[a_{n}k_{n}(\kappa R_{1})P_{n}(\cos \theta_1)+b_{n}\sum_{m=0}^{\infty}(2m+1)B_{nm}(\kappa \mathcal{L})i_{m}(\kappa R_{1})P_{m}(\cos\theta_1)\right]=\sum_{n=0}^{\infty}c_{n}R_{1}^{n}P_{n}(\cos \theta_{1})
\end{equation}\normalsize
associating the common terms in $P_{n}(\cos \theta_1)$, and since the Legendre polynomials are orthogonal \cite{arfken2005} for all $\theta_{1}$, we have
\begin{eqnarray}
a_{j}k_{j}+b_{0}(2j+1)B_{0j}i_{j}+b_{1}(2j+1)B_{1j}i_{j}+\cdots+b_{k}(2j+1)B_{kj}i_{j}+\cdots-c_{j}R_{1}^{j}=0,\mbox{$\forall j\in\mathbb{N}$}\nonumber
\end{eqnarray}
or 
\begin{equation}
c_{j}=\frac{1}{R_{1}^{j}}\alpha_{j}
\label{eq:cj}
\end{equation}
where
\begin{equation}
\alpha_{j}\equiv a_{j}k_{j}(\kappa R_{1})+(2j+1)i_{j}(\kappa R_{1})\sum_{k=0}^{\infty}b_{k}B_{kj}(\kappa \mathcal{L}),\mbox{$\forall j\in\mathbb{N}$}\,.
\label{eq:alphaj}
\end{equation}
The second boundary conditions for sphere $1$ gives
\begin{eqnarray}
\epsilon_{p}\sum_{n=0}^{\infty}c_{n}n\,R_{1}^{n-1}P_{n}(\cos \theta_1)-\epsilon_{m}\sum_{n=0}^{\infty}\left[a_{n}\kappa k^{\prime}_{n}(\kappa R_{1})P_{n}(\cos \theta_1)+\frac{}{}\right.\nonumber\\
\left.+b_{n}\sum_{m=0}^{\infty}(2m+1)B_{nm}(\kappa \mathcal{L})\kappa i^{\prime}_{m}(\kappa R_{1})P_{m}(\cos\theta_1)\right]&=&\frac{\sigma}{\epsilon_{0}}
\end{eqnarray}
where
\begin{eqnarray}
k^{\prime}_{j}\equiv \left.\frac{d}{dx}k_{j}(x)\right|_{x=\kappa R_{1}^{-}}\nonumber\\
i^{\prime}_{j}\equiv \left.\frac{d}{dx}i_{j}(x)\right|_{x=\kappa R_{1}^{+}} \,.
\end{eqnarray}
Regrouping all the common terms in $P_{n}(\cos \theta_{1})$ again, using the Legendre polynomials orthogonality and multiplying both sides of this expression by $R_{1}/\epsilon_{m}$, we find 
\begin{equation}
j\left(\frac{\epsilon_{p}}{\epsilon_{m}}\right)c_{j} R_{1}^{j}-\kappa R_{1}\left[a_{j}k_{j}^{\prime}+(2j+1)i_{j}^{\prime}\sum_{k=0}^{\infty}b_{k}B_{kj}\right]=\frac{R_{1}\sigma}{\epsilon}\delta_{j0} \,.
\end{equation}
Using in the above expression $c_{j}$ given in Eq.(\ref{eq:cj}) we have
\begin{eqnarray}
\left[j\left(\frac{\epsilon_{p}}{\epsilon_{m}}\right) k_{j}(\kappa R_{1})-\kappa R_{1}k_{j}^{\prime}(\kappa R_{1})\right]a_{j}+(2j+1)\cdot\nonumber\\
\cdot\left[j\left(\frac{\epsilon_{p}}{\epsilon_{m}}\right) i_{j}(\kappa R_{1})-\kappa R_{1}i_{j}^{\prime}(\kappa R_{1})\right]\sum_{k=0}^{\infty}B_{kj}(\kappa \mathcal{L})b_{k}&=&\frac{R_{1}\sigma}{\epsilon}\delta_{j0},\mbox{$\forall j\in\mathbb{N}$}
\end{eqnarray}
where $\epsilon\equiv\epsilon_{m}\epsilon_{0}$. Defining
\begin{eqnarray}
A_{j}&\equiv&j\left(\frac{\epsilon_{p}}{\epsilon_{m}}\right) k_{j}(\kappa R_{1})-\kappa R_{1}k_{j}^{\prime}(\kappa R_{1})\nonumber\\
\mathbb{B}_{jk}&\equiv&(2j+1)B_{kj}(\kappa \mathcal{L})\left[j\left(\frac{\epsilon_{p}}{\epsilon_{m}}\right) i_{j}(\kappa R_{1})-\kappa R_{1}i_{j}^{\prime}(\kappa R_{1})\right]
\label{eq:ABcoeff}
\end{eqnarray}
we can finally rewrite the previous equation as  
\begin{equation}
A_{j}a_{j}+\sum_{k=0}^{\infty}\mathbb{B}_{jk}b_{k}=\frac{R_{1}\sigma}{\epsilon}\delta_{j0},\mbox{$\forall j\in\mathbb{N}$ }
\label{eq:sys1} \,.
\end{equation}
%%%%%%%%%%%%%%%%%%%%%%%%%%%%%%%%%%%%%%%%%%%%%%%%%%%%%%%%%%%%%%%%%%%%%%
For the sphere 2, we have similar expressions for the boundary conditions:
\begin{equation}
d_{j}=\frac{1}{R_{2}^{j}}\beta_{j}
\label{eq:dj}
\end{equation} 
with
\begin{equation}
\beta_{j}\equiv b_{j}k_{j}(\kappa R_{2})+(2j+1)i_{j}(\kappa R_{2})\sum_{k=0}^{\infty}a_{k}B_{kj}(\kappa \mathcal{L})
\end{equation}
and
\begin{eqnarray}
(2j+1)\left[j\left(\frac{\epsilon_{p}}{\epsilon_{m}}\right) i_{j}(\kappa R_{2})-\kappa R_{2}i_{j}^{\prime}(\kappa R_{2})\right]\sum_{k=0}^{\infty}B_{kj}a_{k}+\nonumber\\
\left[j\left(\frac{\epsilon_{p}}{\epsilon_{m}}\right) k_{j}(\kappa R_{2})-\kappa R_{2}k_{j}^{\prime}(\kappa R_{2})\right]b_{j}=\frac{R_{2}\sigma}{\epsilon}\delta_{j0} \,.
\end{eqnarray}
Similarly, defining 
\begin{eqnarray}
C_{j}&\equiv&j\left(\frac{\epsilon_{p}}{\epsilon_{m}}\right) k_{j}(\kappa R_{2})-\kappa R_{2}k_{j}^{\prime}(\kappa R_{2})\nonumber\\
\mathbb{D}_{jk}&\equiv&(2j+1)B_{kj}(\kappa \mathcal{L})\left[j\left(\frac{\epsilon_{p}}{\epsilon_{m}}\right) i_{j}(\kappa R_{2})-\kappa R_{2}i_{j}^{\prime}(\kappa R_{2})\right]\nonumber\\
\label{eq:CDcoeff}
\end{eqnarray}
we have
\begin{equation}
\sum_{k=0}^{\infty}\mathbb{D}_{jk}a_{k}+C_{j}b_{j}=\frac{R_{2}\sigma}{\epsilon}\delta_{j0} \,.
\label{eq:sys2}
\end{equation}
The equations Eq.(\ref{eq:sys2}) and Eq.(\ref{eq:sys1}), constitute the linear system which we must solve in order to find the coefficients:
\begin{equation}
\left\{
\begin{array}{ll}
A_{j}a_{j}+\sum_{k=0}^{\infty}\mathbb{B}_{jk}b_{k}=\frac{R_{1}\sigma}{\epsilon}\delta_{j0}&\\
&, \mbox{$\forall j\in\mathbb{N}$}\\
\sum_{k=0}^{\infty}\mathbb{D}_{jk}a_{k}+C_{j}b_{j}=\frac{R_{2}\sigma}{\epsilon}\delta_{j0}&
\end{array}
\right.
\label{eq:lsys5}
\end{equation}
\\
%%%%%%%%%%%%%%%%%%%%%%%%%%%%%%%%%%%%%%%%%%%%%%%%%%%%%%%%%%%%%%%%%%%%%%%%%%%%%%%%%%%%%%%%%%%%
\textbf{Linear System Resolution}
\\
\par Rewriting the Eq.(\ref{eq:sys2}) as
\begin{equation}
b_{k}=\frac{1}{C_{k}}\left(\frac{R_{2}\sigma}{\epsilon}\delta_{k0}-\sum_{l=0}^{\infty}\mathbb{D}_{kl}a_{l}\right), \mbox{$\forall k\in\mathbb{N}$}
\label{eq:sysb}
\end{equation}
we can substitute it in Eq.(\ref{eq:sys1}) to obtain
\begin{equation}
A_{j}a_{j}+\left(\frac{R_{2}\sigma}{\epsilon}\right)\frac{\mathbb{B}_{j0}}{C_{0}}-\sum_{l=0}^{\infty}\sum_{k=0}^{\infty}\frac{\mathbb{B}_{jk}\mathbb{D}_{kl}}{C_{k}}a_{l}=\frac{R_{1}\sigma}{\epsilon}\delta_{j0}, \mbox{$\forall j\in\mathbb{N}$} \,.
\end{equation}
Rearranging the terms, we can rewrite this expression as
\begin{equation}
\sum_{l=0}^{\infty}\mathbb{G}_{jl}a_{l}=E_{j}, \mbox{$\forall j\in\mathbb{N}$}
\label{eq:sysa}
\end{equation}
where
\begin{equation}
\mathbb{G}_{jl}\equiv\mathbb{F}_{jl}-\delta_{jl}A_{l}
\label{eq:Gcoeff}
\end{equation}
and
\begin{eqnarray}
E_{j}&\equiv&\left(\frac{R_{2}\sigma}{\epsilon}\right)\frac{\mathbb{B}_{j0}}{C_{0}}-\left(\frac{R_{1}\sigma}{\epsilon}\right)\delta_{j0}\nonumber\\
\mathbb{F}_{jl}&\equiv&\sum_{k=0}^{\infty}\frac{\mathbb{B}_{jk}\mathbb{D}_{kl}}{C_{k}} \,.\nonumber\\
\label{eq:EFcoeff}
\end{eqnarray}
In matrix notation, Eq.(\ref{eq:sysa}) can be finally rewritten as
\begin{equation}
\mathbf{G}\cdot\mathbf{a}=\mathbf{E} \,.
\label{eq:linsysavec}
\end{equation}
Multiplying both sides by $\mathbf{G}^{-1}$, and using the matrix association law, we have
\begin{eqnarray}
\mathbf{G}^{-1}\cdot\left(\mathbf{G}\cdot\mathbf{a}\right)&=&\left(\mathbf{G}^{-1}\cdot\mathbf{G}\right)\cdot\mathbf{a}=\mathbf{G}^{-1}\cdot\mathbf{E} \,.
\end{eqnarray}
Since $\mathbf{G}^{-1}\cdot\mathbf{G}=\textbf{1}$, we have
\begin{equation}
\mathbf{a}=\mathbf{G}^{-1}\cdot\mathbf{E} \,.
\label{eq:coef}
\end{equation}
Knowing $\mathbf{a}$, we can find $\mathbf{b}$ via Eq.(\ref{eq:sysb}) and finally solve the system. 
%%%%%%%%%%%%%%%%%%%%%%%%%%%%%%%%%%%%%%%%%%%%%%%%%%%%%%%%%%%%%%%%%%%%%%%
%%%%%%%%%%%%%%%%%%%%%%%%%%%%%%%%%%%%%%%%%%%%%%%%%%%%%%%%%%%%%%%%%%%%%%
\\
\section{Double Layer Force Calculation}

\subsubsection{Spherical Surfaces}
\label{sec:maxspher}
\par Following Ref.\cite{Chan94} we will now calculate the force in the sphere 1 due to the surface electrical charge density and the solution electrolytes. According to Fig.(\ref{fig:Maxwell}), if we choose a spherical surface $\partial\Re_{1}$  which embraces sphere's $1$ electrical surface charge density, this force in the Z direction will be [see Appendix A for details] 
\begin{figure}
\centering
\includegraphics*[width=14cm]{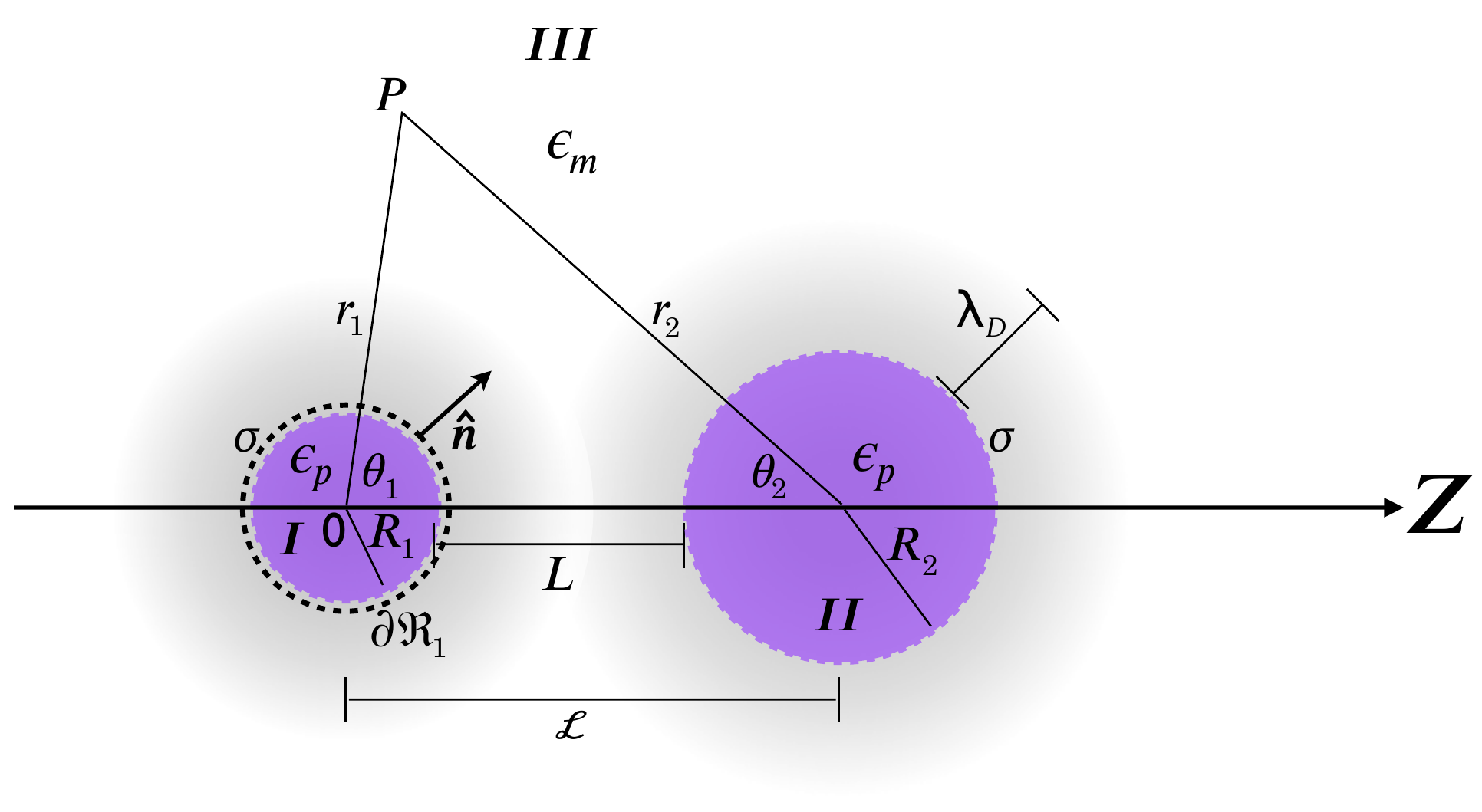}
\caption{Scheme of the spherical surface $\partial\Re_{1}$  embracing sphere's 1 electrical surface charge density $\sigma$ used in double Layer force calculation via Maxwell Stress Tensor.}
\label{fig:Maxwell}
\end{figure}
\begin{eqnarray}
F_{z}=\hat{\textbf{z}}\cdot \textbf{F}_{1}&=&\oint_{\partial\Re_{1}}\hat{\textbf{z}}\cdot\left(\stackrel{\leftrightarrow}{T}-\Pi \textbf{1}\right)\cdot\hat{\textbf{n}}dA
\label{eq:forceon12}
\end{eqnarray}
where 
\begin{eqnarray}
T_{ij}&=&\epsilon\left(E_{i}E_{j}-\frac{1}{2}\delta_{ij}E^{2}\right)\nonumber\\
\Pi&\equiv&\frac{\epsilon\kappa^{2}\psi^{2}}{2}
\label{eq:MaxOsm}
\end{eqnarray} 
are the Maxwell stress tensor and the osmotic pressure respectively. Using the spherical coordinates as in the Figure.(\ref{fig:Maxwell}), and by the axial symmetry we then have
\begin{eqnarray}
\vect{\hat{n}}&=&\vect{\hat{r}}_{1},\nonumber\\
\vect{\hat{z}}&=&\cos\theta_{1}\vect{\hat{r}}_{1}-\sin\theta_{1}\vect{\hat{\theta}}_{1}
\end{eqnarray}
and
\begin{equation}
dA=R^{2}_{1}\sin{\theta_{1}}d\theta_{1}d\phi_{1}\,.
\end{equation} 
As a result
\begin{eqnarray}
F_{z}&=&2\pi R^{2}_{1}\int_{0}^{\pi}\left(\cos\theta_{1}\hat{\textbf{r}}_{1}-\sin\theta_{1}\hat{\theta}_{1}\right)\cdot\left(\stackrel{\leftrightarrow}{T}-\Pi\oper{1}\right)\cdot\hat{\textbf{r}}_{1}\sin{\theta_{1}}d\theta_{1}\,.\nonumber\\
\end{eqnarray} 
Using Eq.(\ref{eq:MaxOsm}), together with the expressions
\begin{eqnarray}
\vect{\hat{r}}_{1}\cdot\overleftrightarrow{T}\cdot\vect{\hat{r}}_{1}&=&(\vect{\hat{r}}_{1}\cdot\overleftrightarrow{T})_{j}(\vect{\hat{r}}_{1})_{j}=r_{1i}T_{ij}r_{1j}=T_{rr}\nonumber\\
&=&\frac{\epsilon}{2}\left(E^{2}_{r}-E^{2}_{\theta}-E^{2}_{\phi}\right)\,,\nonumber\\
\vect{\hat{\theta}}_{1}\cdot\overleftrightarrow{T}\cdot\vect{\hat{r}}_{1}&=&(\vect{\hat{\theta}}_{1}\cdot\overleftrightarrow{T})_{j}(\vect{\hat{r}}_{1})_{j}=\theta_{1i}T_{ij}r_{1j}=T_{\theta r}\nonumber\\
&=&\epsilon\left(E_{\theta}E_{r}\right)\,,\nonumber\\
\vect{\hat{r}}_{1}\cdot\oper{1}\cdot\vect{\hat{r}}_{1}&=&(\vect{\hat{r}}_{1}\cdot\oper{1})_{j}(\vect{\hat{r}}_{1})_{j}=r_{1i}\delta_{ij}r_{1j}=r_{1i}r_{1i}=1\,,\nonumber\\
\vect{\hat{\theta}}_{1}\cdot\oper{1}\cdot\vect{\hat{r}}_{1}&=&(\vect{\hat{\theta}}_{1}\cdot\oper{1})_{j}(\vect{\hat{r}}_{1})_{j}=\theta_{1i}\delta_{ij}r_{1j}=\theta_{1i}r_{1i}=0
\end{eqnarray}
where $i=r,\theta,\phi$, we then have
\begin{equation}
F_{z}=2\pi R^{2}_{1}\epsilon\int_{0}^{\pi}\left[\frac{1}{2}\left(E^{2}_{r}-E^{2}_{\theta}-E^{2}_{\phi}-\kappa^{2}\psi^{2}\right)\cos\theta_{1}-E_{\theta}E_{r}\sin{\theta_{1}}\right]\sin{\theta_{1}}d\theta_{1}\,.
\end{equation}
Using $\mu=\cos\theta_{1}$, we can finally rewrite it as
\begin{equation}
F_{z}=2\pi R^{2}_{1}\epsilon\int_{-1}^{1}\left[\frac{1}{2}\left(E^{2}_{r}-E^{2}_{\theta}-E^{2}_{\phi}-\kappa^{2}\psi^{2}\right)\mu-E_{\theta}E_{r}\left(1-\mu^{2}\right)^{1/2}\right]d\mu
\label{eq:forcezout}
\end{equation}
which is the general desired  expression. It is important to note that all field components are evaluated over the surface $r_{1}=R_{1}^{+}$, that is, in the region $\!I\!I\!I$. 

%%%%%%%%%%%%%%%%%%%%%%%%%%%%%%%%%%%%%%%%%%%%%%%%%%%%%%%%%%%%%%%%%%%%%%%%%%%%%%%%%%%%%%%%%%%%%%%%%%%%%%%%%%%%%%%%%%%%%%%%%%%%%%%%%%%%%%%%%%%%%%%%%%%%%%%%%%%%%%%%%%%%%%%%%%%%%%%%%%%%%%%%%%%%%%%%%%%%%%%%%%%%%%%%%%%%%%%%%%%%%%%%%%%%%%%%%%%%%%%%%%%%%%%%%%%%%%%%%%%%%%%%%%%%%%%%%%%%%%%%%%%%%%%%%%%%%%%%%%%%%%%%%%%%%%%%%%%%%%%%%%%%%%%%%%%%%%%%%%%%%%%%%%%%%%%%%%%%%%%%%%%%%%%%%%%%%%%%%%%%%%%%%%%%%%%%%%%%

%\subsubsection{Boundary Conditions applied to the Force}
Since the outside electric fields calculations are rather cumbersome, we follow Refs.\cite{Chan94,chan93} idea and express the outside electric fields in Eq.(\ref{eq:forcezout}) in terms of the inside electric fields using the boundary conditions Eq.(\ref{eq:bcond}). As a result, for the radial component of the electric field, the second boundary condition gives
\begin{equation}
\left.E^{I\!I\!I}_{r}(r_{1},\theta_{1})\right|_{r_{1}=R_{1}^{+}}=\frac{\sigma}{\epsilon}+\left(\frac{\epsilon_{p}}{\epsilon_{m}}\right)\left.E^{I}_{r}(r_{1},\theta_{1})\right|_{r_{1}=R_{1}^{-}}
\end{equation}
and
\begin{eqnarray}
\left.E^{I\!I\!I}_{\theta}(r_{1},\theta_{1})\right|_{r_{1}=R_{1}^{+}}=\left.E^{I}_{\theta}(r_{1},\theta_{1})\right|_{r_{1}=R_{1}^{-}}\,,\nonumber\\
\left.E^{I\!I\!I}_{\phi}(r_{1},\theta_{1})\right|_{r_{1}=R_{1}^{+}}=\left.E^{I}_{\phi}(r_{1},\theta_{1})\right|_{r_{1}=R_{1}^{-}}\nonumber\\
\end{eqnarray}
for the parallel surface ones \cite{Jackson98}. Furthermore, using the first boundary condition
\begin{equation}
\left.\psi^{I\!I\!I}(r_{1},\theta_{1})\right|_{r_{1}=R_{1}^{+}}=\left.\psi^{I}(r_{1},\theta_{1})\right|_{r_{1}=R_{1}^{-}} \,.
\end{equation}
The electric field inside the sphere 1 can be rewritten as a function of the potential inside as 
\small
\begin{eqnarray}
E^{I}_{r}(r_{1},\theta_{1})&=&-\vect{\hat{r}}_{1}\cdot\left.\nabla\psi^{I}(r_{1},\theta_{1})\right|_{r_{1}=R_{1}^{-}}=-\left.\frac{\partial\psi^{I}}{\partial r_{1}}(r_{1},\theta_{1})\right|_{r_{1}=R_{1}^{-}}\,.\nonumber\\
E^{I}_{\theta}(r_{1},\theta_{1})&=&-\vect{\hat{\theta}}_{1}\cdot\left.\nabla\psi^{I}(r_{1},\theta_{1})\right|_{r_{1}=R_{1}^{-}}=-\frac{1}{r_{1}}\left.\frac{\partial\psi^{I}}{\partial\theta_{1}}(r_{1},\theta_{1})\right|_{r_{1}=R_{1}^{-}}=\frac{1}{r_{1}}\left(1-\mu^{2}\right)^{1/2}\left.\frac{\partial\psi^{I}}{\partial\mu}(r_{1},\theta_{1})\right|_{r_{1}=R_{1}^{-}}\,.\nonumber\\
E^{I}_{\phi}(r_{1},\theta_{1})&=&-\vect{\hat{\phi}}_{1}\cdot\left.\nabla\psi^{I}(r_{1},\theta_{1})\right|_{r_{1}=R_{1}^{-}}=-\frac{1}{r_{1}\sin\theta}\left.\frac{\partial\psi^{I}}{\partial\phi_{1}}(r_{1},\theta_{1})\right|_{r_{1}=R_{1}^{-}}=0 \,\,.\nonumber\\
\end{eqnarray}\normalsize
Using the above expressions, we can rewrite Eq.(\ref{eq:forcezout}) as 
\small
\begin{eqnarray}
F_{z}&=&2\pi R^{2}_{1}\epsilon\int_{-1}^{1}\left\{\frac{1}{2}\left[\left(\frac{\sigma}{\epsilon}+\frac{\epsilon_{p}}{\epsilon_{m}}E^{I}_{r}\right)^{2}-\left(E^{I}_{\theta}\right)^{2}-\kappa^{2}\left(\psi^{I}\right)^{2}\right]\mu-E^{I}_{\theta}\left(\frac{\sigma}{\epsilon}+\frac{\epsilon_{p}}{\epsilon_{m}}E^{I}_{r}\right)\left(1-\mu^{2}\right)^{1/2}\right\}d\mu\nonumber\\
&=&\pi\epsilon\int_{-1}^{1}\left\{\left[R^{2}_{1}\left(\frac{\sigma}{\epsilon}-\frac{\epsilon_{p}}{\epsilon_{m}}\left.\frac{\partial\psi^{I}}{\partial r_{1}}\right|_{r_{1}=R_{1}^{-}}\right)^{2}-\left(1-\mu^{2}\right)\left(\left.\frac{\partial\psi^{I}}{\partial\mu}\right|_{r_{1}=R_{1}^{-}}\right)^2-\left(\kappa R_{1}\right)^{2}\left(\psi^{I}\right)^{2}\right]\mu\right.\nonumber\\
&&\left.-2R_{1}\left.\frac{\partial\psi^{I}}{\partial\mu}\right|_{r_{1}=R_{1}^{-}}\left(\frac{\sigma}{\epsilon}-\frac{\epsilon_{p}}{\epsilon_{m}}\left.\frac{\partial\psi^{I}}{\partial r_{1}}\right|_{r_{1}=R_{1}^{-}}\right)\left(1-\mu^{2}\right)\right\}d\mu\,\,.\nonumber\\
\end{eqnarray}\normalsize
Substituting Eq.(\ref{eq:psiin}) in the above expression, we have
\begin{eqnarray}
F_{z}&=&\pi\epsilon\left[-\left(\frac{4c_{1}R_{1}}{3}\right)\left(\frac{\sigma R_{1}}{\epsilon}\right)\left(\frac{\epsilon_{p}}{\epsilon_{m}}+2\right)+\left(\frac{\epsilon_{p}}{\epsilon_{m}}\right)^{2}\sum_{n=0}^{\infty}nc_{n}R_{1}^{n}\sum_{m=0}^{\infty}mc_{m}R_{1}^{m}\mathbb{C}_{1}(n,m)-\right.\nonumber\\
&&\left.-\sum_{n=0}^{\infty}c_{n}R_{1}^{n}\sum_{m=0}^{\infty}c_{m}R_{1}^{m}\mathbb{C}_{3}(n,m)-\left(\kappa R_{1}\right)^2\sum_{n=0}^{\infty}c_{n}R_{1}^{n}\sum_{m=0}^{\infty}c_{m}R_{1}^{m}\mathbb{C}_{1}(n,m)+\right.\nonumber\\
&&\left.+2\left(\frac{\epsilon_{p}}{\epsilon_{m}}\right)\sum_{n=0}^{\infty}c_{n}R^{n}_{1}\sum_{m=0}^{\infty}mc_{m}R^{m}_{1} \mathbb{C}_{2}(n,m)\right]\nonumber\\
\label{eq:finalF}
\end{eqnarray}
where 
\begin{eqnarray}
\mathbb{C}_{1}(n,m)&=&\int_{-1}^{1}\mu P_{n}(\mu)P_{m}(\mu)d\mu\nonumber\\
&=&\left\{\begin{array}{ll}\frac{2(m+1)}{(2m+3)(2m+1)}, &\mbox{$n=m+1$}\\
\frac{2m}{(2m+1)(2m-1)},&\mbox{$n=m-1$}\\
0,&\mbox{$n\neq m\pm 1$}
\label{eq:integralC1}
\end{array}
\right.
\end{eqnarray}
\begin{eqnarray}
\mathbb{C}_{2}(n,m)&=&\int_{-1}^{1}\left(1-\mu^{2}\right)P_{n}(\mu)P^{\prime}_{m}(\mu)d\mu\nonumber\\
&=&\left\{\begin{array}{ll}-\frac{2m(m+1)}{(2m+3)(2m+1)}, &\mbox{$n=m+1$}\\
\frac{2m(m+1)}{(2m+1)(2m-1)},&\mbox{$n=m-1$}\\
0,&\mbox{$n\neq m\pm 1$}
\end{array}
\right.
\label{eq:integralC2}
\end{eqnarray}
\begin{eqnarray}
\mathbb{C}_{3}(n,m)&=&\int_{-1}^{1}\mu\left(1-\mu^{2}\right)P^{\prime}_{n}(\mu)P^{\prime}_{m}(\mu)d\mu \nonumber\\
&=&\left\{\begin{array}{ll}\frac{2m(m+1)(m+2)}{(2m+3)(2m+1)}, &\mbox{$n=m+1$}\\
\frac{2m(m-1)(m+1)}{(2m+1)(2m-1)},&\mbox{$n=m-1$}\\
0,&\mbox{$n\neq m\pm 1$} \,\,.
\label{eq:integralC3}
\end{array}
\right.
\end{eqnarray}
Using expressions Eq.(\ref{eq:cj}) for the $c_{i}$ coefficients, we can finally rewrite Eq.(\ref{eq:finalF}) in a more adequate form for numerical calculations:
\begin{eqnarray}
F_{z}&=&\pi\epsilon\left[-\left(\frac{4\alpha_{1}}{3}\right)\left(\frac{\sigma R_{1}}{\epsilon}\right)\left(\frac{\epsilon_{p}}{\epsilon_{m}}+2\right)+\left(\frac{\epsilon_{p}}{\epsilon_{m}}\right)^{2}\sum_{n=0}^{\infty}\sum_{m=0}^{\infty}nm\,\alpha_{n}\alpha_{m}\mathbb{C}_{1}(n,m)\right.\nonumber\\
&&\left.-\sum_{n=0}^{\infty}\sum_{m=0}^{\infty}\alpha_{n}\alpha_{m}\mathbb{C}_{3}(n,m)-\left(\kappa R_{1}\right)^2\sum_{n=0}^{\infty}\sum_{m=0}^{\infty}\alpha_{n}\alpha_{m}\mathbb{C}_{1}(n,m)\right.\nonumber\\
&&\left.+2\left(\frac{\epsilon_{p}}{\epsilon_{m}}\right)\sum_{n=0}^{\infty}\sum_{m=0}^{\infty}m\,\alpha_{n}\alpha_{m} \mathbb{C}_{2}(n,m)\right]
\label{eq:finalFalpha}
\end{eqnarray}
where the $\alpha_{j}$ coefficients are given by Eq.(\ref{eq:alphaj}). Analyzing this expression, one would understand the resolution strategy: once the linear system Eq.(\ref{eq:lsys5}) is solved, \textit{i.e.} $a_{i}$ and $b_{i}$ are known, one can find the $c_{i}$ and the $\alpha_{i}$ coefficients via Eq.(\ref{eq:cj}) and Eq.(\ref{eq:alphaj}) respectively, and then the force through the above expression.
\par
Prior to numerically solving this general problem, let us analyze two important particular regimes. 
%%%%%%%%%%%%%%%%%%%%%%%%%%%%%%%%%%%%%%%%%%%%%%%%%%%%%%%%%%%%%%%%%%%%%%%%%%%%%%%%%%%%%%%%%%%%%%%%%%%%%%%%%%%%%%%%%%%%%%%%%%%%%%%%%%%%%%%%%%%%%%%%%%%%%%%%%%%%%%%%%%%%%%%%%%%%%%%%%%%%%%%%%%%%%%%%%%%%%%%%%%%%%%%%%%%%%%%%%%%%%%%%%%%%%%%%%%%%%%%%%%%%%%%%%%%%%%%%%%%%%%%%%%%%%%%%%%%%%%%%%%%%

%%%%%%%%%%%%%%%%%%%%%%%%%%%%%%%%%%%%%%%%%%%%%%%%%%%%%%%%%%%%%%%%%%%%%%%%%%%%%%%%%%%%%%%%%%%%%%%%%%%%%%%%%%%%%%%%%%%%%%%%%%%%%%%%%%%%%%%%%%%%%%%%%%%%%%%%%%%%%%%%%%%%%%%%%%%%%%%%%%%%%%%%%%%%%%%%%%%%%%%%%%%%%%%%%%%%%%%%%%%%%%%%%%%%%%%%%%%%%%%%%%%%%%%%%%%%%%%%%%%%%%%%%%%%%%%%%%%%%%%%%%%%%%%%%%%%%%%%%%%%%%%%%%%%%%%%%%%%%%%%%%%%%%%%%%%%%%%%%%%%%%%%%%%%%%%%%%%%%%%%%%%%%%%%%%%%%%%%%%%%%%%%%%%%%%%%%%%%%%%%%%%%%%%%%%%%%%%%%%%%%%%%%%
\subsection{Limits}
\subsubsection{LSA Limit}
\label{LSA}
\begin{figure}
\centering
\includegraphics*[width=14cm]{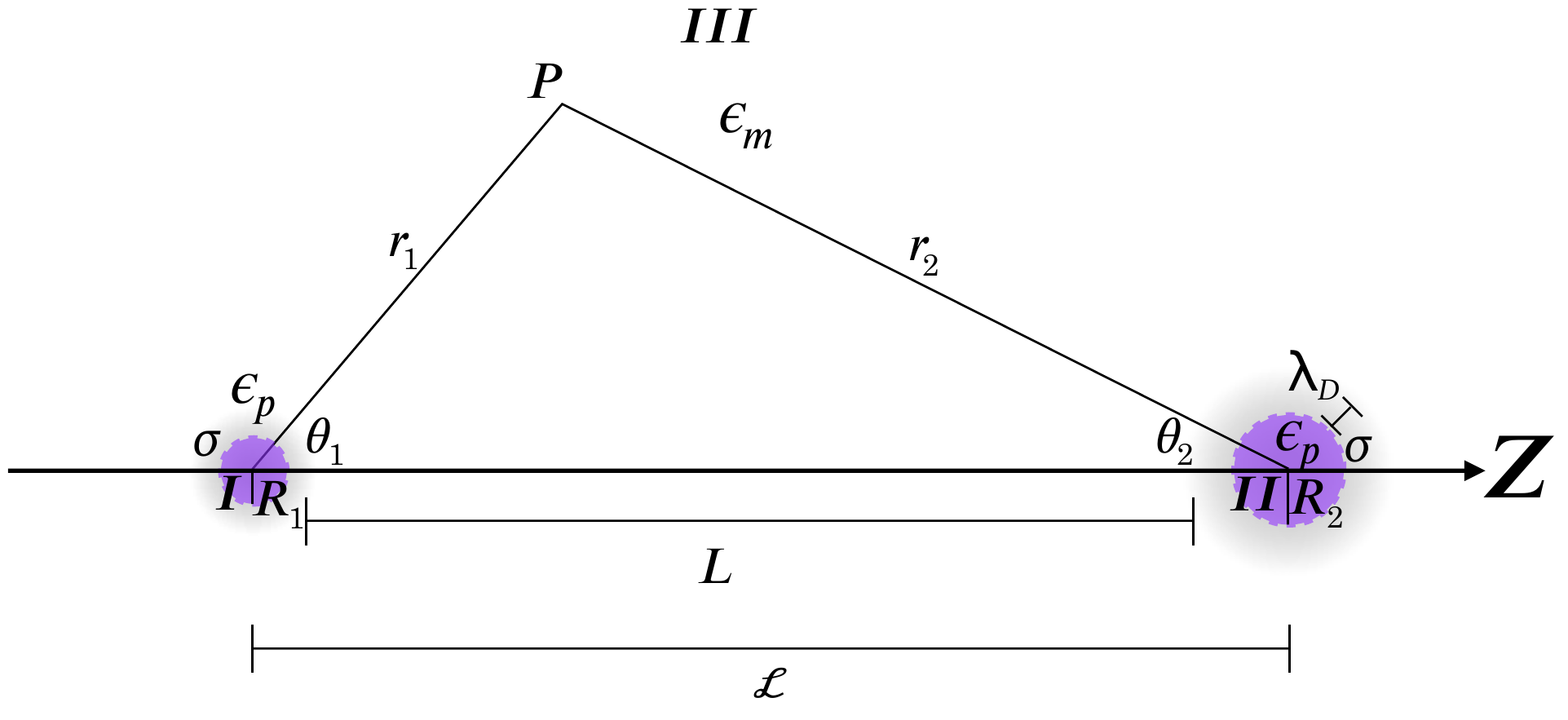}
\caption{Scheme of the LSA approximation regime.}
\label{fig:3}
\end{figure}
\par Let us check the potential and the force expressions in the \textit{linear superposition approximation} (LSA) \cite{bell1970}. Consider now the situation depicted in Fig.\ref{fig:3} in which the two spheres are so far apart that the distance $L$ between them is much greater than the Debye length $\lambda_{D}=1/\kappa$, \emph{i.e.} 
\begin{equation} 
\kappa L\gg1
\label{eq:lsacondL}
\end{equation}
or, equivalently
\begin{equation}
\kappa\mathcal{L}\gg 1+\kappa\left(R_{1}+R_{2}\right) \,.
\label{eq:lsacondLcal}
\end{equation}
In this situation, it is possible to analytically solve the linear system given by Eq.(\ref{eq:lsys5}). According to Ref.\cite{abramowitz64} the modified spherical Bessel function of third kind can be written as 
\begin{equation}
k_{n}(x)=\frac{\pi}{2x}e^{-x}\sum_{k=0}^{n}(n+\frac{1}{2},k)(2x)^{-k}
\label{eq:knexpansion}
\end{equation}
where 
\begin{equation}
(n+\frac{1}{2},k)\equiv\frac{(n+k)!}{k!\Gamma(n-k+1)}\,.
\label{eq:coeffnk}
\end{equation}
The last result allow us to rewrite the $B_{nm}(\kappa  \mathcal{L})$ coefficients given by Eq.(\ref{eq:Bcoef}) as
\begin{eqnarray}
B_{nm}(\kappa \mathcal{L})&=&\frac{\pi}{2\kappa\mathcal{L}}e^{-\kappa\mathcal{L}}\sum_{\nu=0}^{\infty}A_{nm}^{\nu}\times\nonumber\\
&&\sum_{k=0}^{n+m-2\nu}(n+m-2\nu+\frac{1}{2},k)(2\kappa \mathcal{L})^{-k} \,.\nonumber\\
&&
\end{eqnarray}
Using Eq.(\ref{eq:lsacondLcal}), we can retain only the $k=0,1$ terms in this expression, and rewrite it as
\begin{eqnarray}
B_{nm}(\kappa \mathcal{L})&\approx&\frac{\pi}{2\kappa\mathcal{L}}e^{-\kappa\mathcal{L}}\sum_{\nu=0}^{\infty}A_{nm}^{\nu}\left[(n+m-2\nu+\frac{1}{2},0)+\right.\nonumber\\
&&\left.(n+m-2\nu+\frac{1}{2},1)\frac{1}{2\kappa \mathcal{L}}\right] \,.
\label{eq:Bapprox}
\end{eqnarray}
Now let us look to the linear system Eq.(\ref{eq:linsysavec}). Using the above expression, together with Eq.(\ref{eq:ABcoeff}) and Eq.(\ref{eq:CDcoeff}), we can see that the term $\mathbb{F}_{jl}$ in Eq.(\ref{eq:EFcoeff}) is proportional to $e^{-2\kappa  \mathcal{L}}$. Therefore by hypothesis will be neglected:
\begin{eqnarray}
\mathbb{F}_{jl}\propto\mathbb{B}_{jk}\mathbb{D}_{kl}\propto e^{-2\kappa  \mathcal{L}}\rightarrow 0 \,.
\end{eqnarray}
As a result, Eq. (\ref{eq:Gcoeff}) reduces to
\begin{eqnarray}
\mathbb{G}_{jl}\approx-\delta_{jl}A_{l}
\end{eqnarray}
which allows us to write the linear system Eq.(\ref{eq:sysa}) as
\begin{eqnarray}
A_{j}a_{j}=-E_{j}, \mbox{$\forall j\in\mathbb{N}$}
\end{eqnarray}
or
\begin{eqnarray}
a_{j}=-\frac{E_{j}}{A_{j}}=\frac{1}{A_{j}}\left[\left(\frac{R_{1}\sigma}{\epsilon}\right)\delta_{j0}-\left(\frac{R_{2}\sigma}{\epsilon}\right)\frac{\mathbb{B}_{j0}}{C_{0}}\right], \mbox{$\forall j\in\mathbb{N}$}\nonumber\\
\label{eq:acoeflsa}
\end{eqnarray}
where we have used Eq.(\ref{eq:ABcoeff}). Using this result in Eq.(\ref{eq:sysb}) and neglecting again the terms proportional to $e^{-2\kappa  \mathcal{L}}$, we will find
\begin{eqnarray}
b_{j}=\frac{1}{C_{j}}\left[\left(\frac{R_{2}\sigma}{\epsilon}\right)\delta_{j0}-\left(\frac{R_{1}\sigma}{\epsilon}\right)\frac{\mathbb{D}_{j0}}{A_{0}}\right],  \mbox{$\forall j\in\mathbb{N}$} \,.
\label{eq:bcoeflsa}
\end{eqnarray}
Now, if we make a more radical approximation and neglect the terms proportional to $e^{-\kappa  \mathcal{L}}$, the above coefficients will reduce to 
\begin{eqnarray}
\left\{ \begin{array}{ll}
a_{j}=\frac{1}{A_{j}}\left(\frac{R_{1}\sigma}{\epsilon}\right)\delta_{j0}&\nonumber\\
&, \mbox{$\forall j\in\mathbb{N}$}\\
b_{j}=\frac{1}{C_{j}}\left(\frac{R_{2}\sigma}{\epsilon}\right)\delta_{j0}&
\end{array} \right.
\label{eq:abcocer}
\end{eqnarray}
which is the desired linear system solution. Substituting these results in Eq.(\ref{eq:psiout}), we finally have 
\begin{eqnarray}
\psi^{I\!I\!I}(P)= \mbox{C}_{1} \frac{e^{-\kappa r_{1}}}{r_{1}}+\mbox{C}_{2}\frac{e^{-\kappa r_{2}}}{r_{2}}
\label{eq:potentialLSA}
\end{eqnarray}
where 
\begin{eqnarray}
\mbox{C}_{1}&\equiv&\frac{\pi}{2A_{0}}\left(\frac{R_{1}\sigma}{\epsilon}\right)=\psi_{01}R_{1}e^{\kappa R_{1}},\nonumber\\
\mbox{C}_{2}&\equiv&\frac{\pi}{2C_{0}}\left(\frac{R_{2}\sigma}{\epsilon}\right)=\psi_{02}R_{2}e^{\kappa R_{2}}\,.
\label{eq:Cpsi}
\end{eqnarray}
In these expressions
\begin{equation}
\psi_{0i}\equiv\frac{R_{i}}{1+\kappa R_{i}}\left(\frac{\sigma}{\epsilon}\right),i=1,2
\label{eq:psifar}
\end{equation}
are the surface potential of the spheres, $A_{0}=-\kappa R_{1}k_{0}^{\prime}(\kappa R_{1})$, $C_{0}=-\kappa R_{2}k_{0}^{\prime}(\kappa R_{2})$, and we have used $k_{0}(x)=\pi e^{-x}/2x$. From this result, we can see that; in this limit, the spheres are so far apart that the outside potential at a point P is given by the sum of potentials exactly equal to the one obtained in the example (\ref{sec:example}) for one independent sphere. Therefore Eq.(\ref{eq:potentialLSA}) can be written as
\begin{equation}
\psi^{I\!I\!I}(P)=\psi_{1}(\vect{r}_{1})+\psi_{2}(\vect{r}_{2})
\end{equation}
where 
\begin{equation}
\psi_{i}(\vect{r}_{i})\equiv \mbox{C}_{i} \frac{e^{-\kappa r_{i}}}{r_{i}},\,i=1,2\,\,.
\end{equation}
In other words, the electrostatic potential is equal to the addition of the one produced by each independent sphere.
%%%%%%%%%%%%%%%%%%%%%%%%%%%%%%%%%%%%%%%%%%%%%%%%%%%%%%%%%%%%%%%%%%%%%%%%%%%%%%%%%%%%%%%%%%%%%%%%%
\\
\subsubsection{\textbf{Force in LSA}}
\par Performing a lengthy but straightforward calculation (see Appendix B for details), the force expression is given by
\begin{eqnarray}
F_{z}&=&-4\pi\epsilon\left(\frac{\sigma}{\epsilon}\right)^{2}\frac{(1+\kappa\mathcal{L})}{\left(1+\kappa R_{1}\right)\left(1+\kappa R_{2}\right)}\frac{(R_{1}R_{2})^{2}}{\mathcal{L}^{2}}\mathcal{F}(\kappa R_{1})e^{-\kappa\left(\mathcal{L}-R_{1}-R_{2}\right)}\nonumber\\
&=&-4\pi\epsilon\psi_{01}\psi_{02}\left(1+\kappa\mathcal{L}\right)\frac{R_{1}R_{2}}{\mathcal{L}^{2}}\mathcal{F}(\kappa R_{1})e^{-\kappa\left(\mathcal{L}-R_{1}-R_{2}\right)}\nonumber\\
&&
\label{eq:lsaforce}
\end{eqnarray}
where 
\begin{eqnarray}
\mathcal{F}\left(\kappa R_{1}\right)&\equiv&\frac{2+2\kappa R_{1}+\left(\kappa R_{1}\right)^{2}+\left(\kappa R_{1}-1\right)\frac{\epsilon_{p}}{\epsilon_{m}}}{2+2\kappa R_{1}+\left(\kappa R_{1}\right)^{2}+(1+\kappa R_{1})\frac{\epsilon_{p}}{\epsilon_{m}}} \,.\nonumber\\
&&
\end{eqnarray}
In the limit $\epsilon_{p}\rightarrow 0$, $\mathcal{F}(\kappa R_{1})\rightarrow 1$, the above expression reduces exactly to the one given in \cite{bell1970}:
\begin{eqnarray}
F_{z}&\rightarrow&-4\pi\epsilon\left(\frac{\sigma}{\epsilon}\right)^{2}\frac{\left(1+\kappa\mathcal{L}\right)}{\left(1+\kappa R_{1}\right)\left(1+\kappa R_{2}\right)}\frac{(R_{1}R_{2})^{2}}{\mathcal{L}^{2}}e^{-\kappa\left(\mathcal{L}-R_{1}-R_{2}\right)}\nonumber\\
&=&-4\pi\epsilon\psi_{01}\psi_{02}\left(1+\kappa\mathcal{L}\right)\frac{R_{1}R_{2}}{\mathcal{L}^{2}}e^{-\kappa\left(\mathcal{L}-R_{1}-R_{2}\right)}\,.\nonumber\\
&&
\end{eqnarray}
From this result, we can see that if the spheres have the same adsorbed surface charge densities, the force in sphere 1 in the LSA limit will always be negative, \textit{i.e.} opposite to the Z-axis orientation, which means that it is a repulsive force. Additionally, note that the exponential argument depends only on $L=\mathcal{L}-R_{1}-R_{2}$, which is the smallest distance between the spheres.

%%%%%%%%%%%%%%%%%%%%%%%%%%%%%%%%%%%%%%%%%%%%%%%%%%%%%%%%%%%%%%%%%%%%%%%%%%%%%%%%%%%%%%%%

\subsubsection{Proximity Force Approximation (PFA) Limit}
\label{PFA}
\begin{figure}
\centering
\includegraphics*[width=12cm]{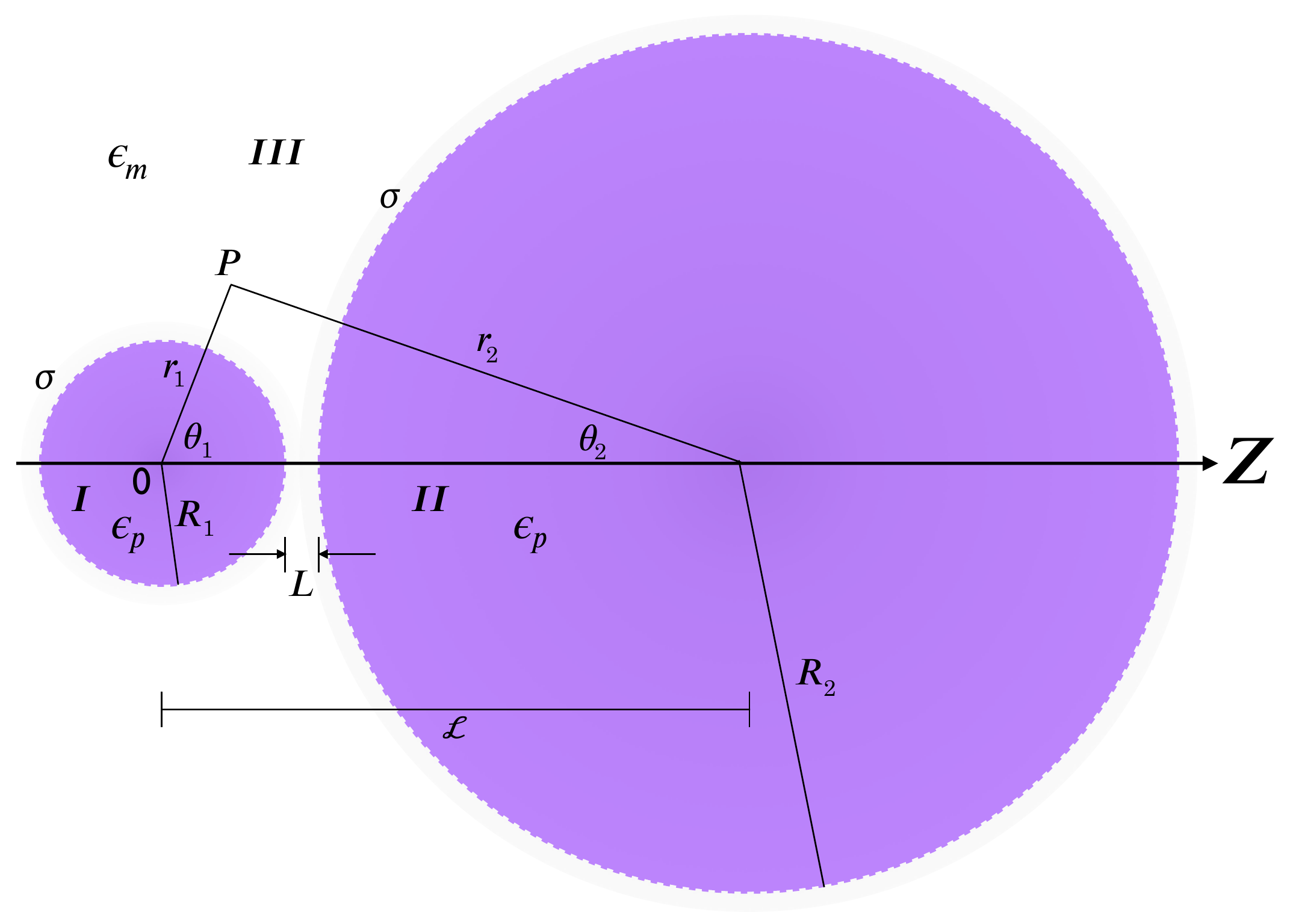}
\caption{Scheme of the PFA approximation regime.}
\label{fig:4}
\end{figure}
Another very important limit to the force between spheres is the proximity force approximation (PFA) regime or Deryaguin approximation (\cite{Butt, israelachvili11}). PFA limit is when the interaction range between the spheres, given by the Debye length $\lambda_{D}=\frac{1}{\kappa}$, and the smaller distance $L$, are much smaller than the minor of their radius (Fig. \ref{fig:4}):
\begin{eqnarray}
\frac{1}{\kappa}&\ll& R_{1},\nonumber\\
L&\ll&R_{1}
\end{eqnarray}
where we suppose $R_{1}<R_{2}$. To calculate the total force on sphere 1 in this limit, let us calculate first the potential energy between two dielectric half regions 1 and 2 which are separated by a distance $\mbox{z}$, with electrical surface charge density in its plane surfaces, and have relative permittivities $\epsilon_{p}$ (Fig.\ref{fig:planes}). Again, $z:z$ electrolytes are dissolved in the region between them and are in thermal equilibrium with a thermal bath, being $n_{0}$ its bulk concentration.
\begin{figure}[h]
\centering
\includegraphics*[width=11cm]{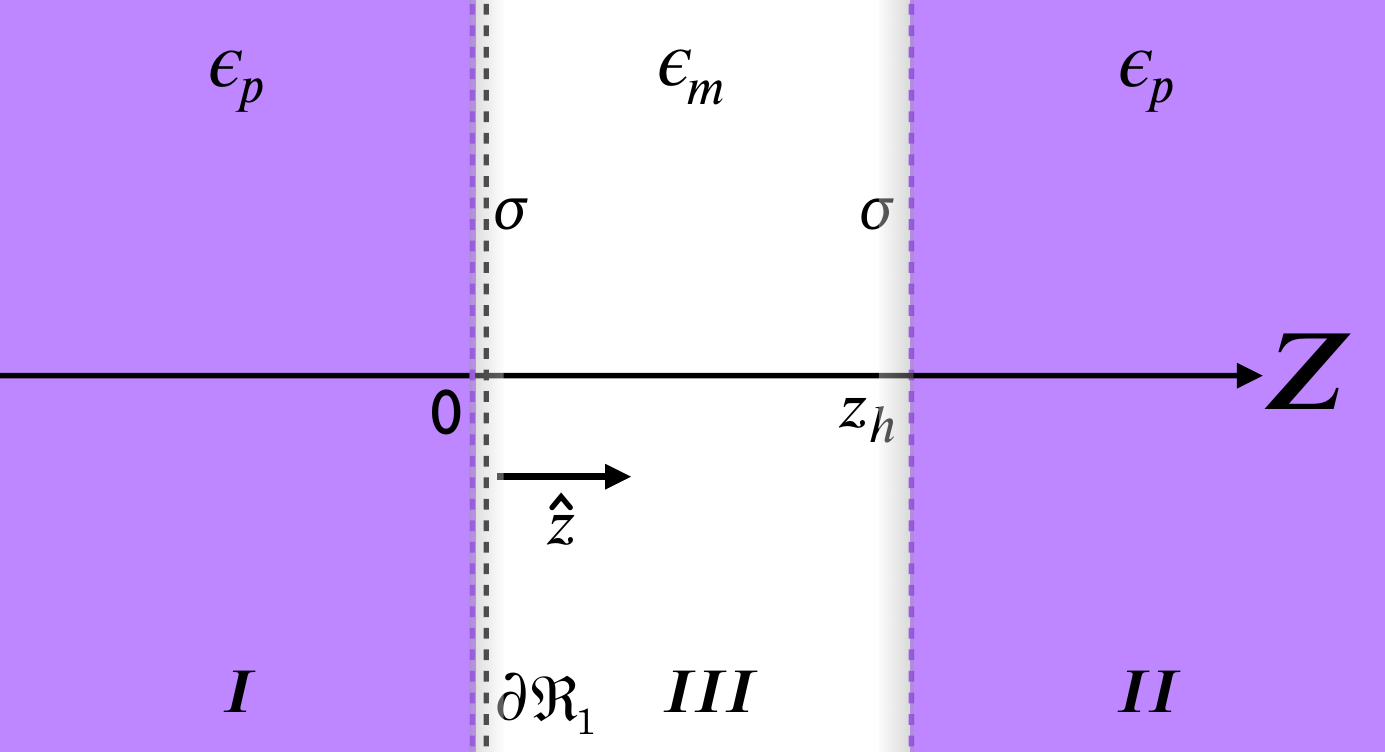}
\caption{Scheme of two dielectric half regions 1 and 2 which are separated by a distance $\mbox{z}$, with electrical surface charge density $\sigma$ in its surfaces, and relative permittivities $\epsilon_{p}$.}
\label{fig:planes}
\end{figure}
\par
\textit{(a) Boundary Value Problem}. For this configuration, we must solve the equations
\par
\begin{eqnarray}
\frac{d^{2}\psi^{I\!I\!I}}{dz^{2}}(z)&=&\kappa^{2}\psi^{I\!I\!I}(z)\,\,\,\,\mbox{(if $0\leq z \leq z_{h}$)}\nonumber\\
\frac{d^{2}\psi^{I}}{dz^{2}}(z)&=&0\,\,\,\,\mbox{(if $z<0$)}\nonumber\\
\frac{d^{2}\psi^{I\!I}}{dz^{2}}(z)&=&0\,\,\,\,\mbox{(if $z>z_{h}$)}
\end{eqnarray}
subject to the boundary conditions 
\begin{eqnarray}
\left.\psi^{I}(z)\right|_{z=0^{-}}=\left.\psi^{I\!I\!I}(z)\right|_{z=0^{+}}&&\nonumber\\
\epsilon_{p}\left.\frac{d\psi^{I}}{dz}(z)\right|_{z=0^{-}}-\epsilon_{m}\left.\frac{d\psi^{I\!I\!I}}{dz}(z)\right|_{z=0^{+}}&=&\frac{\sigma}{\epsilon_{0}}\nonumber\\
&&
\end{eqnarray}
for the $z=0$ plane and 
\begin{eqnarray}
\left.\psi^{I\!I\!I}(z)\right|_{z=z^{-}_{h}}=\left.\psi^{I\!I}(z)\right|_{z=z^{+}_{h}}&&\nonumber\\
\epsilon_{m}\left.\frac{d\psi^{I\!I\!I}}{dz}(z)\right|_{z=z^{-}_{h}}-\epsilon_{p}\left.\frac{d\psi^{I\!I}}{dz}(z)\right|_{z=z^{+}_{h}}&=&\frac{\sigma}{\epsilon_{0}}\nonumber\\
&&
\end{eqnarray}
for the $z=z_{h}$ plane. Additionally, $\psi^{I}(z)\rightarrow constant$ for $z\rightarrow -\infty$ and $\psi^{I\!I}(z)\rightarrow constant$ for $z\rightarrow \infty$.
\par
\textit{(b) General Solution and Boundary Conditions}. 
The general solution for the potential outside and inside the half regions are respectively given by
\begin{eqnarray}
\psi^{I\!I\!I}(z)&=&\mbox{A}e^{-\kappa z}+\mbox{B}e^{\kappa z},\,\,\,0\leq z \leq z_{h}\nonumber\\
\psi^{I}(z)&=&\mbox{C}\,\,\,z<0\nonumber\\
\psi^{I\!I}(z)&=&\mbox{D},\,\,\,z>z_{h} \,.
\label{eq:planesol}
\end{eqnarray}
Substituting Eq.(\ref{eq:planesol}) in these equations, we have, for $z=0$,
\begin{eqnarray}
&(i)&\mbox{C}=\mbox{A}+\mbox{B}\nonumber\\
&(ii)&\kappa\epsilon_{m}\left(\mbox{A}-\mbox{B}\right)=\frac{\sigma}{\epsilon_{0}}\nonumber\\
&&
\end{eqnarray}
and, for $z=z_{h}$,
\begin{eqnarray}
&(iii)&\mbox{A}e^{-\kappa z_{p}}+\mbox{B}e^{\kappa z_{p}}=D\nonumber\\
&(iv)&\kappa\epsilon_{m}\left(\mbox{B}e^{\kappa z_{h}}-\mbox{A}e^{-\kappa z_{h}}\right)=\frac{\sigma}{\epsilon_{0}} \,.\nonumber\\
&&
\end{eqnarray}
Grouping these equations, we then have
\begin{eqnarray}
&(i)&\mbox{A}+\mbox{B}-\mbox{C}+0.\mbox{D}=0\nonumber\\
&(ii)&\mbox{A}-\mbox{B}+0.\mbox{C}+0.\mbox{D}=\frac{\sigma}{\kappa\epsilon_{m}\epsilon_{0}}\nonumber\\
&(iii)&\beta\mbox{A}+\alpha\mbox{B}+0.\mbox{C}-\mbox{D}=0\nonumber\\
&(iv)&-\beta\mbox{A}+\alpha\mbox{B}+0.\mbox{C}+0.\mbox{D}=\frac{\sigma}{\kappa\epsilon_{m}\epsilon_{0}}\nonumber\\
&&
\end{eqnarray}
where $\alpha\equiv e^{\kappa z_{h}}$ and $\beta\equiv e^{-\kappa z_{h}}=1/\alpha$. Solving this linear system for the $\mbox{A}$ and $\mbox{B}$ coefficients, we then have
\begin{eqnarray}
\Delta&=&\left|
\begin{array}{cccc}
1&1&-1&0\\
1&-1&0&0\\
\beta&\alpha&0&-1\\
-\beta&\alpha&0&0\\
\end{array}\right|\nonumber\\
&=&-2\sinh(\kappa z_{h})\nonumber\\
&&
\end{eqnarray}
\begin{eqnarray}
A&=&\frac{1}{\Delta}\left|
\begin{array}{cccc}
0&1&-1&0\\
\sigma/\kappa\epsilon_{m}\epsilon_{0}&-1&0&0\\
0&\alpha&0&-1\\
\sigma/\kappa\epsilon_{m}\epsilon_{0}&\alpha&0&0\\
\end{array}\right|\nonumber\\
&=&\frac{\sigma}{\kappa\epsilon_{m}\epsilon_{0}}e^{\kappa z_{h}/2}\frac{\cosh(\kappa z_{h}/2)}{\sinh(\kappa z_{h})}
\label{eq:planeA}
\end{eqnarray}
\begin{eqnarray}
B&=&\frac{1}{\Delta}\left|
\begin{array}{cccc}
1&0&-1&0\\
1&\sigma/\kappa\epsilon_{m}\epsilon_{0}&0&0\\
\beta&0&0&-1\\
-\beta&\sigma/\kappa\epsilon_{m}\epsilon_{0}&0&0\\
\end{array}\right|\nonumber\\
&=&\frac{\sigma}{\kappa\epsilon_{m}\epsilon_{0}}e^{-\kappa z_{h}/2}\frac{\cosh(\kappa z_{h}/2)}{\sinh(\kappa z_{h})} \,.\nonumber\\
&&
\label{eq:planeB}
\end{eqnarray}
\par
\textit{(c) Pressure, Potential Energy and Force}. According to Eq.(\ref{eq:forceon12}), the pressure in the surface $\partial\mathcal{R}_{1}$ due to the adsorbed electrical charges in the half region $I\!I$ surface and the medium $I\!I\!I$ electrolytes is given by 
\begin{eqnarray}
\left.P\right|_{z=0^{+}}&=&\vect{\hat{z}}\cdot\left(\overleftrightarrow{T}-\Pi\oper{1}\right)\cdot\vect{\hat{z}}\nonumber\\
&=&(\vect{\hat{z}}\cdot\overleftrightarrow{T})_{j}(\vect{\hat{z}})_{j}-\Pi(\vect{\hat{z}}\cdot\oper{1})_{j}(\vect{\hat{z}})_{j}\nonumber\\
&=&z_{i}T_{ij}z_{j}-\Pi z_{i}\delta_{ij}z_{j}=T_{zz}-\Pi\nonumber\\
&=&\epsilon\left(E^{2}_{z}-\frac{1}{2}E^{2}\right)-\Pi\nonumber\\
&=&\epsilon\left[E^{2}_{z}-\frac{1}{2}\left(E^{2}_{x}+E^{2}_{y}+E^{2}_{z}\right)\right]-\frac{\epsilon\kappa^{2}}{2}\psi^{2}\nonumber\\
&=&\frac{\epsilon}{2}E^{2}_{z}-\frac{\epsilon\kappa^{2}}{2}\psi^{2}\nonumber\\
&=&\frac{\epsilon}{2}\left(\frac{d\psi}{dz}\right)^{2}-\frac{\epsilon\kappa^{2}}{2}\psi^{2}\nonumber\\
&&
\end{eqnarray} 
where we omitted the index $I\!I\!I\!$ for simplicity and we used the fact that electrostatic potential $\psi$ depends only on $\mbox{z}$ in the fifth line of the previous equation. Using Eq.(\ref{eq:planesol}) together with Eq.(\ref{eq:planeA}) and Eq.(\ref{eq:planeB}) in the corresponding result, we finally have  
\begin{eqnarray}
P&=&-2\epsilon\kappa^{2}\mbox{A}\mbox{B}\nonumber\\
&=&-\frac{2\sigma^{2}}{\epsilon_{m}\epsilon_{0}}\frac{\cosh^{2}(\kappa z_{h}/2)}{\sinh^{2}(\kappa z_{h})} \,.\nonumber\\
\label{eq:planepressure}
\end{eqnarray}
\begin{figure}[h]
\centering
\includegraphics*[width=12cm]{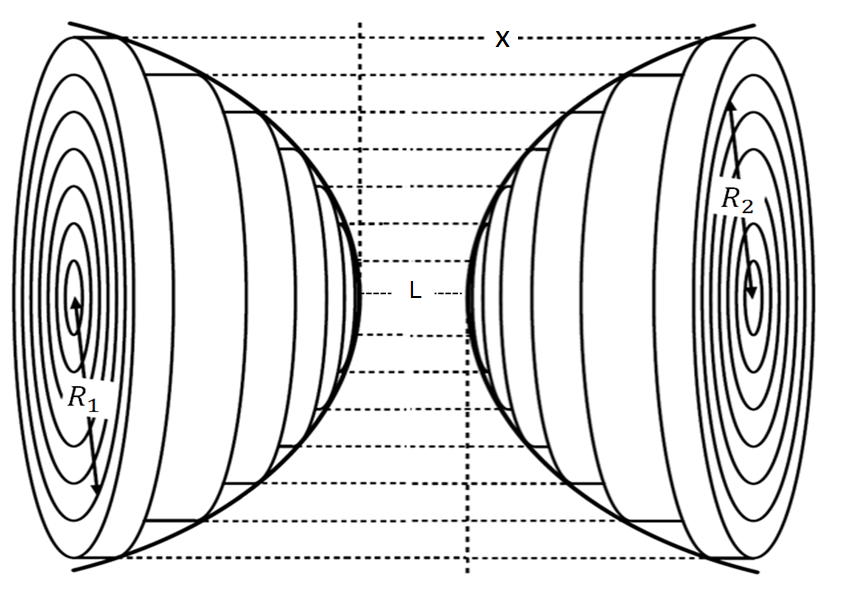}
\caption{Scheme of the capes construction for the spheres in the PFA model. Modification of Fig.(2.5) in \cite{Butt}.}
\label{fig:modibutt}
\end{figure}
With this result we can calculate the interaction potential energy per unit area between the half regions $I$ and $I\!I$ \cite{Butt}:
\begin{equation}
u_{planes}(L)=-\int_{\infty}^{L}P\left(z_{h}\right)dz_{h} \,.
\end{equation}
Using Eq.(\ref{eq:planepressure}), we then have
\begin{eqnarray}
u_{planes}(L)&=&\frac{2\sigma^{2}}{\epsilon_{m}\epsilon_{0}}\int_{\infty}^{L}\frac{\cosh^{2}(\kappa z_{h}/2)}{\sinh^{2}(\kappa z_{h})}dz_{h}\nonumber\\
&=&-\frac{2\sigma^{2}}{\kappa\epsilon_{m}\epsilon_{0}}\frac{1+e^{-\kappa L}}{e^{\kappa L}-e^{-\kappa L}} \,.
\label{eq:energyplanes}
\end{eqnarray}
Once the potential energy per unit area between the two planar surfaces is calculated, we can use the proximity force (or Derjaguin) approximation to calculate the force between any two bodies. This approximation states that the range of interaction between the two curved surfaces (proportional to the Debye length) is proportional to their effective radius, whereby the inverse effective radius is the arithmetic mean of the inverse curvature radii of the surfaces of the involved bodies \cite{rentsch2006}. In PFA the construction of the potential energy between two bodies with minimum separation $L$ can be done through the integration of the potential energy per unit area between two planar surfaces $u_{planes}(x)$ \cite{Butt} with the area restricted to the cross-sectional area of the bodies
\begin{eqnarray}
u_{curves}(L)=\int_{L}^{\infty}u_{planes}\left(x\right)dA
\label{eq:energycurves}
\end{eqnarray}
where $x$ is the separation between the planar surfaces that compose the bodies. For two spheres of different radii we have rotational symmetric configuration (Fig. \ref{fig:modibutt}). If we admit that the effective radius is much smaller than the spheres radii and, therefore, the range of the interaction includes only contributions of the outer caps of the two spheres \cite{Butt, rentsch2006}, the Eq.(\ref{eq:energycurves}) can be written as
\begin{eqnarray}
u_{spheres}(L)=2\pi\left(\frac{R_{1}R_{2}}{R_{1}+R_{2}}\right)\int_{L}^{\infty}u_{planes}\left(x\right)dx\,.
\label{eq:energyspheres}
\end{eqnarray}
Since the force can be calculated as 
\begin{eqnarray}
F(L)=-\frac{du}{dL}=2\pi\left(\frac{R_{1}R_{2}}{R_{1}+R_{2}}\right)u_{planes}(L)
\label{eq:forcesph}
\end{eqnarray}
using Eq.(\ref{eq:energyplanes}), we finally conclude
\begin{eqnarray}
F(L)&=&-\frac{4\pi\sigma^{2}}{\kappa\epsilon_{m}\epsilon_{0}}\left(\frac{R_{1}R_{2}}{R_{1}+R_{2}}\right)\frac{1+e^{-\kappa L}}{e^{\kappa L}-e^{-\kappa L}}\nonumber\\
&&
\end{eqnarray}
which is the textbook expression in Ref.(\cite{Butt}) for the case in which both spheres have the same electrical surface charge density $\sigma$.
\end{chapter}

%% file: chapter4.tex
\begin{chapter}{Numerical Analysis of the LPB Solution}
\newcommand{\scalprod}[2]{\ensuremath{\left\langle{#1}|{#2}\right\rangle}}
\newcommand{\brm}[1]{\boldsymbol{\mathrm{#1}}}
\newcommand{\bsy}[1]{\boldsymbol{{#1}}}
\newcommand{\vect}[1]{\boldsymbol{#1}}
\newcommand{\oper}[1]{\boldsymbol{\mathsf{#1}}}
\newcommand{\vac}{\textsc{vac}}
\newcommand{\sinc}{\ensuremath{\mathrm{sinc}}}
\newcommand{\steve}[1]{{\color{blue} #1}}

\label{cap4}
\hspace{5 mm}
This chapter is about the implementation of a code in the \textit{Mathematica} platform that allows us to obtain and analyze numerical values for the double layer force between two colloidal spheres of different radii.
\\
\section{Definition of the Functions}

\par According to Eq.(\ref{eq:coef}), we have first compute the matrices $\mathbf{G}$ and $\mathbf{E}$ that enable us to find the vector column $\mathbf{a}$, containing the coefficients $a_{n}$. After calculating those coefficients, we find the $b_{n}$ coefficients via Eq.(\ref{eq:sysb}), $c_{i}$ and $\alpha_{i}$ coefficients via Eq.(\ref{eq:cj}) and Eq.(\ref{eq:alphaj}) respectively, and the force from Eq.(\ref{eq:finalFalpha}). We briefly discuss below how we implement the calculation of the matrix $\mathbf{G}$ that leads to the derivation of the double-layer force. 
 \par
 The modified Bessel functions of the first and third kinds $I_{\nu}(x)$ and $K_{\nu}(x)$ are already defined in \textit{Mathematica}. We build From them, the modified \textit{spherical} Bessel functions and their derivatives, following the 
definitions and using results from \cite{abramowitz64}. In the next step, we compute the function $B_{nm}(\kappa \mathcal{L})$ defined by Eq.(\ref{eq:Bcoef}), which is associated to the Bessel Addition Theorem. Here arises the first important question: what is the optimal upper limit $\nu$ for the sum defining $B_{nm}$? 
We have studied numerically the convergence of $B_{nm}$ and have  observed that for a given order in $n$, the maximum upper limit of the sum is $\nu=n$, that is for $\nu>n$, the contribution to $B_{nm}$ in Eq.(\ref{eq:Bcoef}) vanishes. Consequently, the function can be written as
\begin{equation}
B_{nm}(x)=\sum_{\nu=0}^{n}A_{nm}^{\nu}k_{n+m-2\nu}(x)
\end{equation} 
with $x$ representing the separation between the sphere centers in units of the Debye length $\lambda_{D}$: $x\equiv\kappa \mathcal{L}= \mathcal{L}/\lambda_{D}$.

We continue by computing the auxiliary column vector $A_{j}$ and matrix $\mathbb{B}_{jk}$ associated to the sphere $1$ and defined by Eq.(\ref{eq:ABcoeff}). They involve not only the previous quantities, but also the radii of the spheres as well as the ratio between the relative permittivity $\epsilon_{p}$ of the spheres and the medium (solution) $\epsilon_{m}$.
In an analogous way, we compute the auxiliary vector and matrix associated to sphere $2$: $C_{j}$ and $\mathbb{D}_{jk}$ defined by Eq.~(\ref{eq:CDcoeff}).
The next step is to build the matrix function $\mathbb{F}_{jl}$ given by Eq.(\ref{eq:EFcoeff}), which together with the coefficients $A_{j}$ allows us to compute the matrix $\mathbf{G}$. 
At this point,  we also need to set the upper limit for the sum involved in the definition of the matrix $\mathbb{F}_{jl},$ which we define as $\upsilon$: 
\begin{eqnarray}
\mathbb{F}_{jl}&\equiv&\sum_{k=0}^{\upsilon}\frac{\mathbb{B}_{jk}\mathbb{D}_{kl}}{C_{k}}
\label{eq:Fcoeff}
\end{eqnarray}

Once the matrix $\mathbf{G}$ is known and after the evaluation of the $E_{j}$ vector, \textit{Mathematica} can solve the linear system (\ref{eq:coef}) for the coefficients $a_{n}$ giving the potential outside the spheres. Next, we find the $b_{n}$ coefficients using Eq.(\ref{eq:sysb}). We compute  the functions $\alpha_{j}$ Eq.(\ref{eq:alphaj}), $\mathbb{C}_{1}$,  $\mathbb{C}_{2}$, $\mathbb{C}_{3}$ in Eq.(\ref{eq:integralC1}), and finally the expression for the force between the spheres as given by Eq.(\ref{eq:finalFalpha}). In Table \ref{tab:fixed}, we summarize some of the parameters employed in the calculation. In addition to the parameters shown in this table, we take for the spheres surface charge density,  the value $\sigma=10 \mu C/m^{2}$  found when surface polystyrene films are charged at room temperature \cite{ohara1984}. The additional parameters are the following: $N$ represent the maximum multipole order expansion and therefore the quantity of coefficients $a_n$ and $b_n$ (thus the resulting  dimension is $N+1$ and $(N+1)\times(N+1)$ for the square matrices)\footnote{$N=0$ calculates the force with order zero coefficients $a_{0}$ and $b_{0}$ only.}, $L$ is the closest separation between the sphere surfaces, $\upsilon$ previously defined as the upper limit for the matrix $\mathbb{F}_{jl}$ calculation, and the Debye length $\lambda_{D}=1/\kappa$ which is the most important parameter, playing the leading role for the definition of the LSA and PFA limits. We take $\lambda_{D}=100nm$ as a typical value for a water solution. 
\begin{table}[width=11cm]
	\centering
		\begin{tabular}{ ||c|c|c|c|c|| } \hline $R_{1} (\mu m)$ & $R_{2} (\mu m)$ & $\epsilon_{m}$  & $\epsilon_{p}$ & $\epsilon_{0} (C^{2}\,N^{-1}\,m^{-2})$\\ 7.18 & 1.50 & 78.5 & 2.5 & $8.85\times10^{-12}$ \\ \hline \end{tabular}
	\caption{Fixed values used in the numerical calculation. They are: Spheres radii $R_{i}$. Relative permittivity of the medium $\epsilon_{m}$ \cite{Butt}, of the spheres $\epsilon_{p}$ \cite{Butt}, and vacuum permittivity $\epsilon_{0}$ \cite{Jackson98}.}
	\label{tab:fixed}
\end{table}
%%%%%%%%%%%%%%%%%%%%%%%%%%%%%%%%%%%%%%%%%%%%%%%%%%%%%%%%%%%%%%%%%%%%%%%%%%%%%%%%%%%%%%%%%%%%%%%%%%%%%%%%%%%%%%%%%%%%%%%%
\section{Testing the Code}
\par
We start testing the code by comparing it with the LSA limit $\kappa \mathcal{L}\gg 1.$
 We first compare the zeroth order coefficients $a_0$ and $b_0$ with the LSA analytical expressions Eqs.(\ref{eq:acoeflsa},\ref{eq:bcoeflsa}), where terms proportional to $e^{-\kappa  \mathcal{L}}$ can be neglected. 
 For simplicity, we take $N=0$ when using our code. 
  For $\lambda_{D}=100nm$ and using the values given in Table.\ref{tab:fixed}, 
the analytical LSA expressions yield $a_{0}=1.55113\times 10^{30}$ and $b_{0}=68331.0$. 
On the other hand, the exact result for the coefficients depends on the ratio between the separation and the Debye length $\frac{L}{\lambda_{D}}$. In Fig.\ref{fig:coef}, we plot the ratio between the exact and LSA coefficients as a function of distance.
  We can see how the exact coefficients come near to the LSA ones as the separation increases. The LSA accuracies at $L/\lambda_{D}=1$ are $99.999\%$ and $98.5\%$ for $a_{0}$ and $b_{0},$ respectively, while for $L\geq 8\lambda_D$ the accuracy is $(100-1\times10^{-13})\%$ for both coefficients.
\begin{figure}[H]
\centering
\includegraphics*[width=15cm]{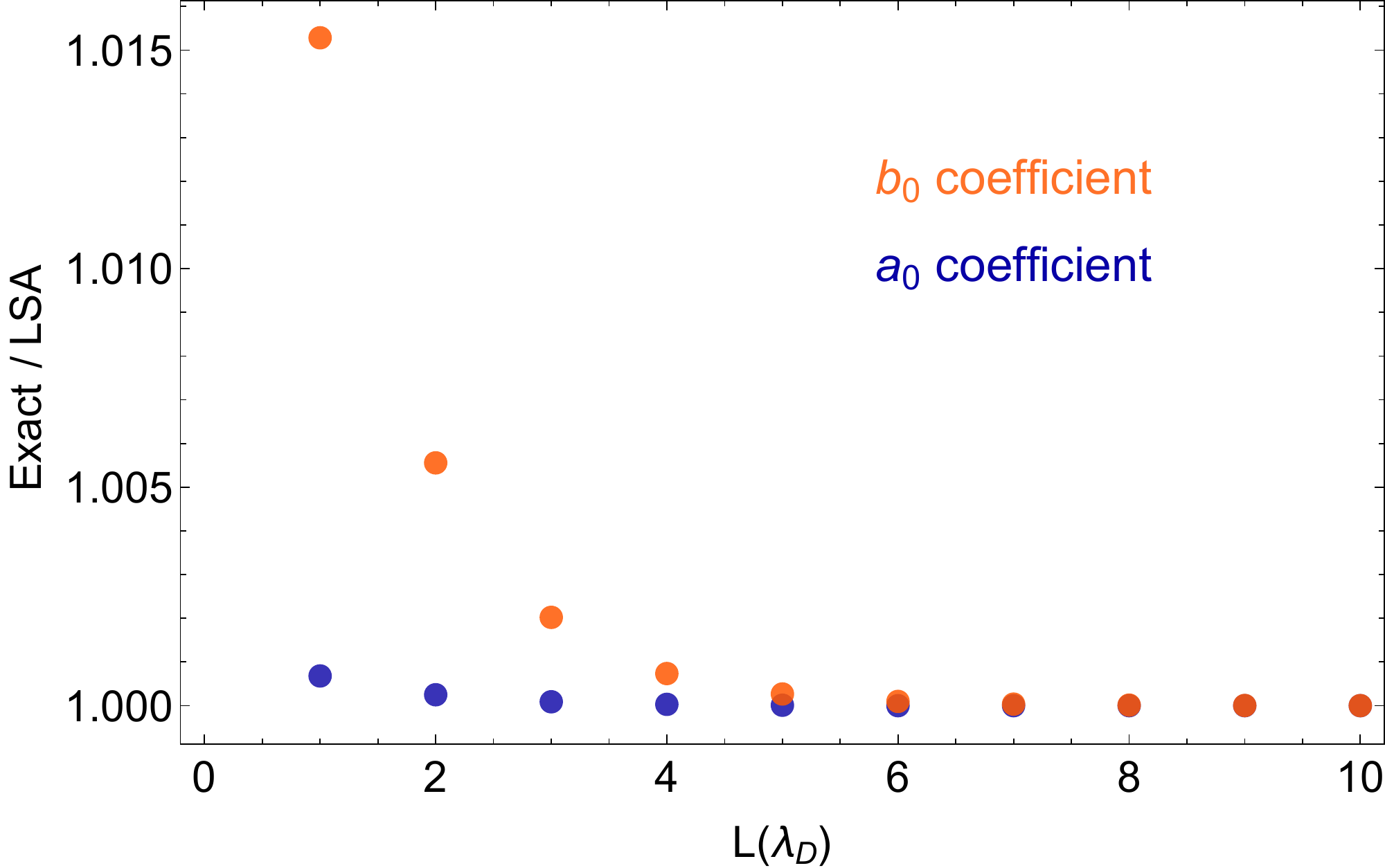}
\caption{Ratio of the coefficients $a_{0}$ and $b_{0}$ of the exact (with $N=0$) and LSA calculations as a function of the edge spheres separation in units of Debye Lengths ($\lambda_{D}=100nm$).}
\label{fig:coef}
\end{figure}

After this initial validation of the code, we tackle the question for the required upper limit $\upsilon$ in the evaluation of the matrix $\mathbb{F}_{jl}$. We calculate and analyze the $a_{n}$ coefficients calculated with different values of 
 $\upsilon$, and as a function of the sphere separation in the range  $10nm\leq L\leq 100nm=\lambda_{D}$. 
We verify that (i) changing the $\upsilon$ values becomes relevant only when the spheres are closer than ($2\lambda_D$); 
(ii) the variations in $\upsilon$ are increasingly relevant for higher order coefficients; 
(iii) given a fixed sphere separation, variations in $\upsilon$ are only relevant until $\upsilon=N/2$. 
With these observations, it is now possible to analyze the force. It is  important to note that the calculation time increases with the quantity of coefficients used in the calculation of the force. Thus, it is of great importance to establish the order $N$ required for numerical convergence. 
Therefore, we calculated the force values for a different fixed closest sphere separation as a function of the matrix order. Hence, we see that the force values have a faster convergence for greater values of separation. Figure.\ref{fig:convergence} shows for two different separations the force variation as a function of the  calculation order $N$. The force for $L=400nm$ has already converged \textit{i.e.} the variation in the order of $N$ does not give rise to significant variations, while for $L=100nm$, although the variations in the force are smaller when $N$ increase, we have not seen convergence yet. However, for $L=200nm$, the force does not change appreciably for $N>5$.
\begin{figure}[H]
\centering
\includegraphics*[width=15cm]{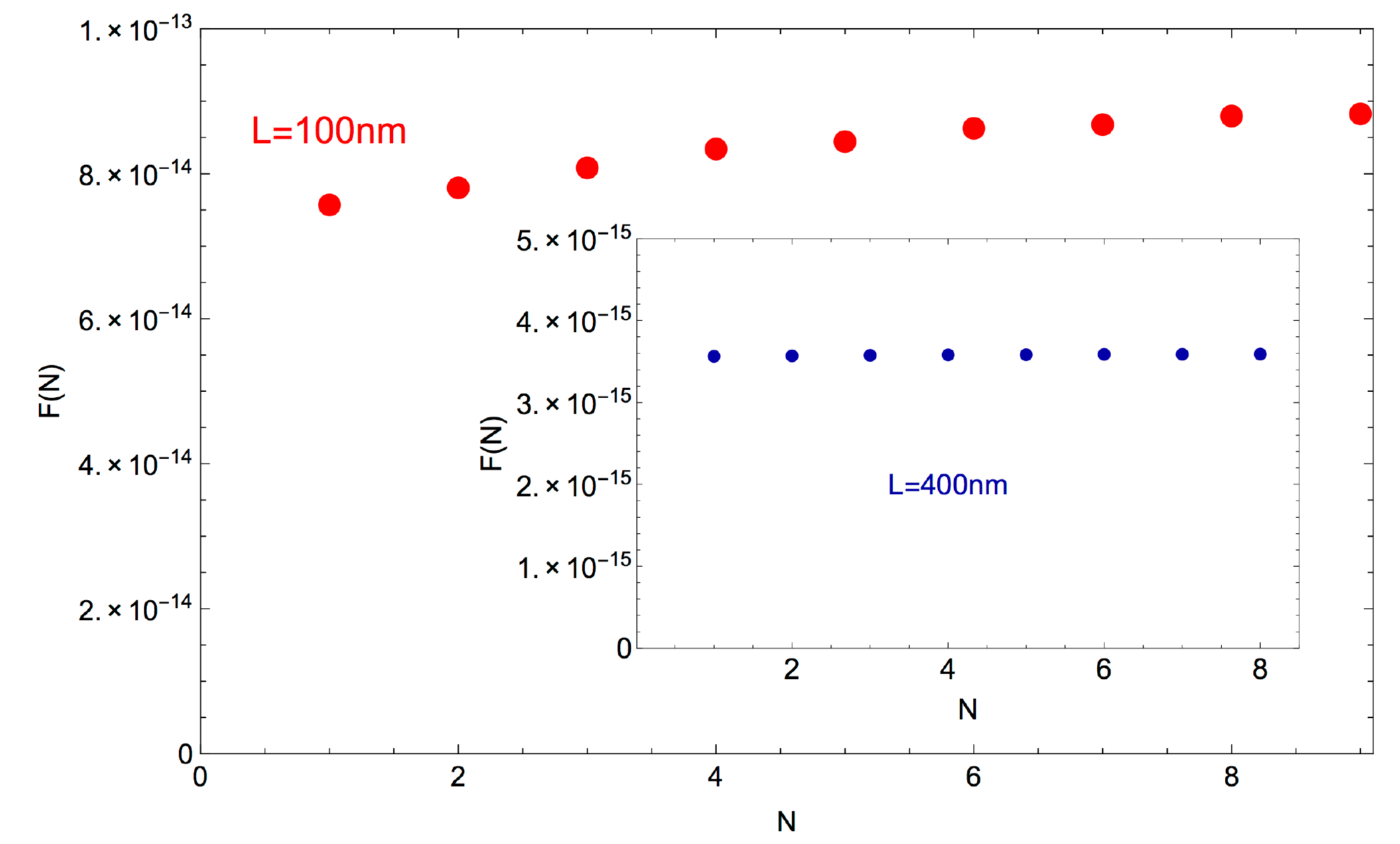}
\caption{Variation of the force for $2$ different fixed closest sphere separation as a function of the multipole expansion order $N$.}
\label{fig:convergence}
\end{figure}
Finally, we compared our exact solution with the analytical limits LSA and PFA. In Figure.\ref{fig:ForceLSAPFA} we demonstrate how the increment in the order $N$ of the calculation becomes relevant with the spheres approach. For separations greater than $200nm$ ($2\lambda_{D}$), there is not a relevant variation of the force values for the different range order $1\leq N\leq10$, in agreement with the LSA limit. For separations smaller than $200nm$, the tendency of the force is deviates from the LSA curve when the spheres approaching is increased, showing that in this region the LSA limit has an underestimation. In this region where the spheres are considered to be near, the PFA limit overestimate the force. However, the tendency is the force approximate to the PFA curve with the increasing of the order of calculation. 
\begin{figure}[H]
\includegraphics*[width=15cm]{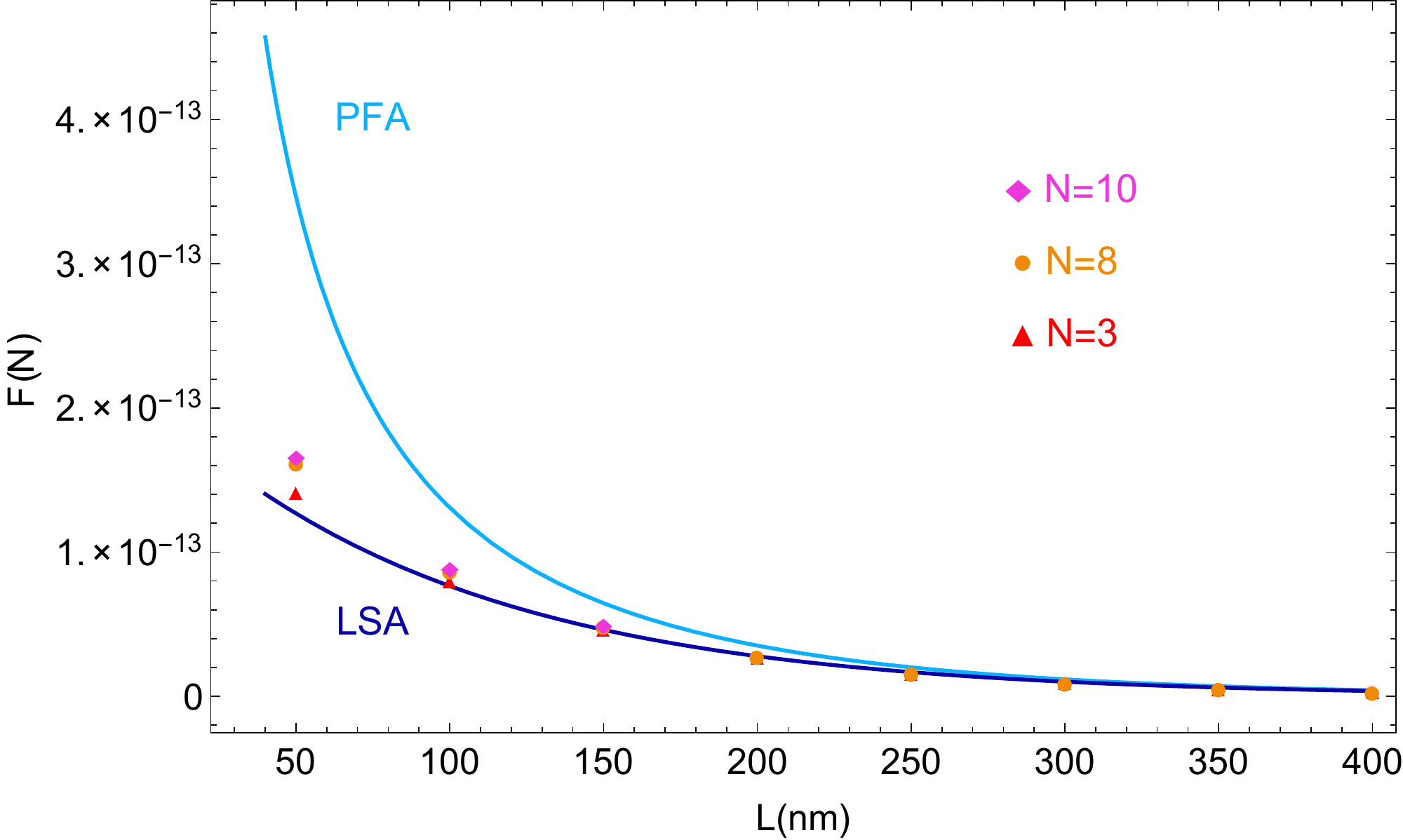}
\caption{Force as a function of the sphere closest separation for the analytical LSA and PFA limits, and three different matrix order $N=3,8,10$.}
\label{fig:ForceLSAPFA}
\end{figure}
\end{chapter}

%% file: chapter5.tex
\begin{chapter}{Conclusions and Perspectives}
\label{cap5}

\hspace{5 mm} 

We have implemented a code in the platform \textit{Mathematica} in order to analyze the analytical solution in multipole expansion for the force between two dissimilar colloidal spheres with uniform and fixed surface charged immersed in a solution of $z:z$ type. The calculation used as maximum order of the expansion $N_{max}=10$ rendered reasonable results in comparison with the LSA and PFA analytical limits. The analysis with the code allowed us to fix the extremal optimal values for some parameter; study the convergence of the exact force values for different separations; orders of the multipole expansions; and finally to compare with the analytical models for the different limits LSA and PFA. For our model, we found that higher order for the force calculation was only relevant for smaller separations than the Debye length when PFA conditions are satisfied. In regions where the LSA model governs ($L\geq3\lambda_{D}$), the accuracy of the code is excellent even for low orders $N\leq 5$.
We will improve the code in order to reduce the calculation time (Timing $T$) even more than $\sqrt{T}$, making further possible force calculations in regions of even closer sphere approach ($L\leq \lambda_{D}/2$) with multipole orders of $N\geq100$. This results will be important in the experiments concerning Casimir and Electrostatic forces, which are being done at the UFRJ Optical Tweezers Laboratory (LPO - COPPEA).
Finally, we are also working in the development of analytical expressions and numerical calculations for the interaction between dielectric and conductor colloidal spheres. Such results will also be important in future LPO - COPPEA experiments, where the force between these type of spheres will be measured.
\end{chapter}

%% file: apendicea.tex
\begin{chapter}{Appendix A: Force Expression}
\label{apendicea}
\newcommand{\scalprod}[2]{\ensuremath{\left\langle{#1}|{#2}\right\rangle}}
\newcommand{\brm}[1]{\boldsymbol{\mathrm{#1}}}
\newcommand{\bsy}[1]{\boldsymbol{{#1}}}
\newcommand{\vect}[1]{\boldsymbol{#1}}
\newcommand{\oper}[1]{\boldsymbol{\mathsf{#1}}}
\newcommand{\vac}{\textsc{vac}}
\newcommand{\sinc}{\ensuremath{\mathrm{sinc}}}
\newcommand{\steve}[1]{{\color{blue} #1}}
\hspace{5 mm} 
\hspace{5 mm} 
\par Let us rewrite Eq.(\ref{eq:PBlO}) in region $I\!I\!I\!$ as
\begin{equation}
\nabla^{2}\psi-\kappa^{2}\psi=0
\end{equation}
where we omit the index $I\!I\!I\!$ for simplicity. Multiplying both sides by $\epsilon\nabla\psi$, we have
\begin{equation}
\epsilon\nabla\psi\nabla^{2}\psi-\epsilon\kappa^{2}\nabla\psi\psi=0
\end{equation}
which can be written in component form
\begin{eqnarray}
\epsilon e_{i}\left(\partial_{i}\psi\right)\left(\partial_{j}\partial_{j}\psi\right)-\epsilon\kappa^{2}e_{i}\partial_{i}\left(\psi\right)\psi&=&0\nonumber\\
\epsilon e_{i}\left[\partial_{j}\left(\partial_{i}\psi\partial_{j}\psi\right)-\left(\partial_{j}\partial_{i}\psi\right)\partial_{j}\psi \right]-\frac{ \epsilon\kappa^{2}}{2}e_{i}\partial_{i}\left(\psi^{2}\right)&=&0\nonumber\\
\epsilon e_{i}\left[\partial_{j}\left(\partial_{i}\psi\partial_{j}\psi\right)-\frac{1}{2}\partial_{i}\left(\partial_{j}\psi\partial_{j}\psi\right) \right]-\frac{\epsilon\kappa^{2}}{2}e_{i}\partial_{i}\left(\psi^{2}\right)&=&0\nonumber\\
\epsilon e_{i}\left[\partial_{j}\left(\partial_{i}\psi\partial_{j}\psi\right)-\frac{1}{2}\delta_{ij}\partial_{j}\left(\partial_{k}\psi\partial_{k}\psi\right) \right]-\frac{\epsilon\kappa^{2}}{2}e_{i}\partial_{i}\left(\psi^{2}\right)&=&0\nonumber\\
 e_{i}\partial_{j}\left\{\epsilon \left[\partial_{i}\psi\partial_{j}\psi-\frac{1}{2}\delta_{ij}\left(\partial_{k}\psi\partial_{k}\psi\right) \right]\right\}-\frac{\epsilon\kappa^{2}}{2}e_{i}\partial_{i}\left(\psi^{2}\right)&=&0\nonumber\\
e_{i}\left\{\partial_{j}\left\{\epsilon \left[\partial_{i}\psi\partial_{j}\psi-\frac{1}{2}\delta_{ij}\left(\partial_{k}\psi\partial_{k}\psi\right) \right]\right\}-\frac{\epsilon\kappa^{2}}{2}\partial_{i}\left(\psi^{2}\right)\right\}&=&0\nonumber\\
\end{eqnarray}
Given that the last equation must be valid for all components, we have
\begin{eqnarray}
\partial_{j}\left\{\epsilon \left[\partial_{i}\psi\partial_{j}\psi-\frac{1}{2}\delta_{ij}\left(\partial_{k}\psi\partial_{k}\psi\right) \right]\right\}-\frac{\epsilon\kappa^{2}}{2}\partial_{i}\left(\psi^{2}\right)&=&0,\nonumber\\
\partial_{j}\left\{\epsilon \left[\partial_{i}\psi\partial_{j}\psi-\frac{1}{2}\delta_{ij}\left(\partial_{k}\psi\partial_{k}\psi\right) \right]\right\}-\frac{\epsilon\kappa^{2}}{2}\delta_{ij}\partial_{j}\left(\psi^{2}\right)&=&0,\nonumber\\
\partial_{j}\left\{\epsilon \left[\partial_{i}\psi\partial_{j}\psi-\frac{1}{2}\delta_{ij}\left(\partial_{k}\psi\partial_{k}\psi\right) \right]-\delta_{ij}\left(\frac{\epsilon\kappa^{2}}{2}\psi^{2}\right)\right\}&=&0,\nonumber\\
\end{eqnarray}
Remembering that $E_{i}=-\partial_{i}\psi$, we get
\begin{equation}
\partial_{j}\left\{\epsilon \left[E_{i}E_{j}-\frac{1}{2}\delta_{ij}\left(E^{2}\right) \right]-\delta_{ij}\left(\frac{\epsilon\kappa^{2}}{2}\psi^{2}\right)\right\}=0
\label{eq:div}
\end{equation}
According to Ref.\cite{griffiths99}, we identify in in the last equation the Maxwell stress tensor
\begin{equation}
T_{ij}=\epsilon\left(E_{i}E_{j}-\frac{1}{2}\delta_{ij}E^{2}\right)
\label{eq:maxwellstresstensor}
\end{equation} 
The other term in Eq.(\ref{eq:div}) is known as the osmotic pressure, which has a purely entropic origin for molecules obeying Boltzmann statistics \cite{israelachvili11}. In fact, defining
\begin{equation}
\Pi\equiv\frac{\epsilon\kappa^{2}\psi^{2}}{2}
\label{eq:osmotic}
\end{equation}
we can rewrite Eq.(\ref{eq:div}) as
\begin{equation}
\partial_{j}\left(T_{ij}-\delta_{ij}\Pi\right)=0
\end{equation}
or
\begin{equation}
\nabla\cdot\left(\overleftrightarrow{T}-\Pi\oper{1}
\right)=0
\label{eq:tensor}
\end{equation}
The conserved quantity $\overleftrightarrow{T}-\Pi\oper{1}$ is related to the total pressure on a surface and allows us to calculate the force on sphere 1 due to the sphere 2 surface electrical charge density and the solution electrolytes.
\end{chapter}

%% file: apendiceb.tex
\begin{chapter}{Appendix B: Force in LSA Approximation}
\label{apendiceb}
\newcommand{\scalprod}[2]{\ensuremath{\left\langle{#1}|{#2}\right\rangle}}
\newcommand{\brm}[1]{\boldsymbol{\mathrm{#1}}}
\newcommand{\bsy}[1]{\boldsymbol{{#1}}}
\newcommand{\vect}[1]{\boldsymbol{#1}}
\newcommand{\oper}[1]{\boldsymbol{\mathsf{#1}}}
\newcommand{\vac}{\textsc{vac}}
\newcommand{\sinc}{\ensuremath{\mathrm{sinc}}}
\newcommand{\steve}[1]{{\color{blue} #1}}
\hspace{5 mm} 

To calculate the force on sphere 1 due to the sphere 2 electrical surface charge density plus the solution electrolytes in the LSA regime, we approximate the involved functions in the same spirit as for the electrostatic potential calculation. In fact, substituting Eq.(\ref{eq:acoeflsa}) and Eq.(\ref{eq:bcoeflsa}) in Eq.(\ref{eq:alphaj}) and neglecting the term $\sim\mathbb{D}_{k0}B_{kj}$, which, again, is proportional to $e^{-2\kappa  \mathcal{L}}$, we have
\footnotesize
\begin{equation}
\alpha_{j}\approx\frac{k_{j}}{A_{j}}\left[\left(\frac{R_{1}\sigma}{\epsilon}\right)\delta_{j0}-\left(\frac{R_{2}\sigma}{\epsilon}\right)\frac{\mathbb{B}_{j0}}{C_{0}}\right]+\left(\frac{R_{2}\sigma}{\epsilon}\right)(2j+1)i_{j}\frac{B_{0j}}{C_{0}}
\end{equation}\normalsize
Now, we must substitute the above result in Eq.(\ref{eq:finalFalpha}) and analyze all the terms separately, with the help of Legendre integrals Eq.(\ref{eq:integralC1}), Eq.(\ref{eq:integralC2}), and Eq.(\ref{eq:integralC3}):
\footnotesize
\begin{eqnarray}
\left(\frac{\epsilon_{p}}{\epsilon_{m}}\right)^{2}\sum_{n=0}^{\infty}\sum_{m=0}^{\infty}nm\alpha_{n}\alpha_{m}\mathbb{C}_{1}(n,m)\approx\left(\frac{\epsilon_{p}}{\epsilon_{m}}\right)^{2}\sum_{n=0}^{\infty}\sum_{m=0}^{\infty}nm\mathbb{C}_{1}(n,m)\times\nonumber&&\\
\left\{\frac{k_{n}}{A_{n}}\left[\left(\frac{R_{1}\sigma}{\epsilon}\right)\delta_{n0}-\left(\frac{R_{2}\sigma}{\epsilon}\right)\frac{\mathbb{B}_{n0}}{C_{0}}\right]+(2n+1)i_{n}\left(\frac{R_{2}\sigma}{\epsilon}\right)\frac{B_{0n}}{C_{0}}\right\}\times\nonumber&&\\
\left\{\frac{k_{m}}{A_{m}}\left[\left(\frac{R_{1}\sigma}{\epsilon}\right)\delta_{m0}-\left(\frac{R_{2}\sigma}{\epsilon}\right)\frac{\mathbb{B}_{m0}}{C_{0}}\right]+(2m+1)i_{m}\left(\frac{R_{2}\sigma}{\epsilon}\right)\frac{B_{0m}}{C_{0}}\right\}=0\nonumber&&\\
&&
\end{eqnarray}
\normalsize
since all the terms with kronecker's delta are zero - there are $n$ and $m$ inside the sum - and we neglect higher terms proportional to $e^{-2\kappa  \mathcal{L}}$.
\footnotesize
\begin{eqnarray}
&&\sum_{n=0}^{\infty}\sum_{m=0}^{\infty}\alpha_{n}\alpha_{m}\mathbb{C}_{3}(n,m)\approx\sum_{n=0}^{\infty}\sum_{m=0}^{\infty}\mathbb{C}_{3}(n,m)\times\nonumber\\
&&\left\{\frac{k_{n}}{A_{n}}\left[\left(\frac{R_{1}\sigma}{\epsilon}\right)\delta_{n0}-\left(\frac{R_{2}\sigma}{\epsilon}\right)\frac{\mathbb{B}_{n0}}{C_{0}}\right]+(2n+1)i_{n}\left(\frac{R_{2}\sigma}{\epsilon}\right)\frac{B_{0n}}{C_{0}}\right\}\times\nonumber\\
&&\left\{\frac{k_{m}}{A_{m}}\left[\left(\frac{R_{1}\sigma}{\epsilon}\right)\delta_{m0}-\left(\frac{R_{2}\sigma}{\epsilon}\right)\frac{\mathbb{B}_{m0}}{C_{0}}\right]+(2m+1)i_{m}\left(\frac{R_{2}\sigma}{\epsilon}\right)\frac{B_{0m}}{C_{0}}\right\}\nonumber\\
&=&\mathbb{C}_{3}(0,0)\left(\frac{k_{0}}{A_{0}}\right)^{2}R^{2}_{1}\left(\frac{\sigma}{\epsilon}\right)^{2}+\frac{k_{0}}{A_{0}C_{0}}R_{1}R_{2}\left(\frac{\sigma}{\epsilon}\right)^{2}\sum_{m=0}^{\infty}\mathbb{C}_{3}(0,m)\left[(2m+1)i_{m}B_{0m}-\frac{k_{m}}{A_{m}}\mathbb{B}_{m0}\right]\nonumber\\
&&+\frac{k_{0}}{A_{0}C_{0}}R_{1}R_{2}\left(\frac{\sigma}{\epsilon}\right)^{2}\sum_{n=0}^{\infty}\mathbb{C}_{3}(n,0)\left[(2n+1)i_{n}B_{0n}-\frac{k_{n}}{A_{n}}\mathbb{B}_{n0}\right]=0\nonumber\\
&&
\end{eqnarray}
\normalsize
since, according to Eq.(\ref{eq:integralC3}), $\mathbb{C}_{3}(0,0)=\mathbb{C}_{3}(0,m)=\mathbb{C}_{3}(n,0)=0$.
\footnotesize
\begin{eqnarray}
&&2\left(\frac{\epsilon_{p}}{\epsilon_{m}}\right)\sum_{n=0}^{\infty}\sum_{m=0}^{\infty}m\alpha_{n}\alpha_{m} \mathbb{C}_{2}(n,m)
\approx2\left(\frac{\epsilon_{p}}{\epsilon_{m}}\right)\sum_{n=0}^{\infty}\sum_{m=0}^{\infty}m\,\mathbb{C}_{2}(n,m)\times\nonumber\\
&&\left\{\frac{k_{n}}{A_{n}}\left[\left(\frac{R_{1}\sigma}{\epsilon}\right)\delta_{n0}-\left(\frac{R_{2}\sigma}{\epsilon}\right)\frac{\mathbb{B}_{n0}}{C_{0}}\right]+(2n+1)i_{n}\left(\frac{R_{2}\sigma}{\epsilon}\right)\frac{B_{0n}}{C_{0}}\right\}\times\nonumber\\
&&\left\{\frac{k_{m}}{A_{m}}\left[\left(\frac{R_{1}\sigma}{\epsilon}\right)\delta_{m0}-\left(\frac{R_{2}\sigma}{\epsilon}\right)\frac{\mathbb{B}_{m0}}{C_{0}}\right]+(2m+1)i_{m}\left(\frac{R_{2}\sigma}{\epsilon}\right)\frac{B_{0m}}{C_{0}}\right\}\nonumber\\
&=&2\left(\frac{\epsilon_{p}}{\epsilon_{m}}\right)\frac{k_{0}}{A_{0}C_{0}}R_{1}R_{2}\left(\frac{\sigma}{\epsilon}\right)^{2}\sum_{m=0}^{\infty}m\,\mathbb{C}_{2}(0,m)\left[(2m+1)i_{m}B_{0m}-\frac{k_{m}}{A_{m}}\mathbb{B}_{m0}\right]\nonumber\\
&=&\frac{8}{3}\left(\frac{\epsilon_{p}}{\epsilon_{m}}\right)\frac{k_{0}}{A_{0}C_{0}}R_{1}R_{2}\left(\frac{\sigma}{\epsilon}\right)^{2}\left(3i_{1}B_{01}-\frac{k_{1}}{A_{1}}\mathbb{B}_{10}\right)
\end{eqnarray}\normalsize
where in the last line we have used the fact that
\footnotesize
\begin{eqnarray}
\mathbb{C}_{2}(0,m)&=&\left\{\begin{array}{ll}
\frac{4}{3} &\mbox{$m=1$}\\
0,&\mbox{$m\neq 1$}
\end{array}
\right.
\end{eqnarray}
\begin{eqnarray}
-\left(\frac{4\alpha_{1}}{3}\right)\left(\frac{\sigma R_{1}}{\epsilon}\right)\left(\frac{\epsilon_{p}}{\epsilon_{m}}+2\right)&\approx&-\frac{4}{3}\left(\frac{\epsilon_{p}}{\epsilon_{m}}+2\right)\frac{1}{C_{0}}R_{1}R_{2}\left(\frac{\sigma}{\epsilon}\right)^{2}\left(3i_{1}B_{01}-\frac{k_{1}}{A_{1}}\mathbb{B}_{10}\right)\nonumber\\
\end{eqnarray}\normalsize
The last term
\footnotesize
\begin{eqnarray}
&&-\left(\kappa R_{1}\right)^2\sum_{n=0}^{\infty}\sum_{m=0}^{\infty}\alpha_{n}\alpha_{m}\mathbb{C}_{1}(n,m)\approx-\left(\kappa R_{1}\right)^2\sum_{n=0}^{\infty}\sum_{m=0}^{\infty}\mathbb{C}_{1}(n,m)\times\nonumber\\
&&\left\{\frac{k_{n}}{A_{n}}\left[\left(\frac{R_{1}\sigma}{\epsilon}\right)\delta_{n0}-\left(\frac{R_{2}\sigma}{\epsilon}\right)\frac{\mathbb{B}_{n0}}{C_{0}}\right]+(2n+1)i_{n}\left(\frac{R_{2}\sigma}{\epsilon}\right)\frac{B_{0n}}{C_{0}}\right\}\times\nonumber\\
&&\left\{\frac{k_{m}}{A_{m}}\left[\left(\frac{R_{1}\sigma}{\epsilon}\right)\delta_{m0}-\left(\frac{R_{2}\sigma}{\epsilon}\right)\frac{\mathbb{B}_{m0}}{C_{0}}\right]+(2m+1)i_{m}\left(\frac{R_{2}\sigma}{\epsilon}\right)\frac{B_{0m}}{C_{0}}\right\}\nonumber\\
&=&-\left(\kappa R_{1}\right)^2\left\{\mathbb{C}_{1}(0,0)\left(\frac{k_{0}}{A_{0}}\right)^{2}R^{2}_{1}\left(\frac{\sigma}{\epsilon}\right)^{2}\right.\nonumber\\
&&+\left.\frac{k_{0}}{A_{0}C_{0}}R_{1}R_{2}\left(\frac{\sigma}{\epsilon}\right)^{2}\sum_{m=0}^{\infty}\mathbb{C}_{1}(0,m)\left[(2m+1)i_{m}B_{0m}-\frac{k_{m}}{A_{m}}\mathbb{B}_{m0}\right]\right.\nonumber\\
&&+\left.\frac{k_{0}}{A_{0}C_{0}}R_{1}R_{2}\left(\frac{\sigma}{\epsilon}\right)^{2}\sum_{n=0}^{\infty}\mathbb{C}_{1}(n,0)\left[(2n+1)i_{n}B_{0n}-\frac{k_{n}}{A_{n}}\mathbb{B}_{n0}\right]\right\}\nonumber\\
&=&-\frac{4}{3}\left(\kappa R_{1}\right)^2\frac{k_{0}}{A_{0}C_{0}}R_{1}R_{2}\left(\frac{\sigma}{\epsilon}\right)^{2}\left(3i_{1}B_{01}-\frac{k_{1}}{A_{1}}\mathbb{B}_{10}\right)
\end{eqnarray}\normalsize
where we have used $\mathbb{C}_{1}(0,0)=0$,
\footnotesize
\begin{eqnarray}
\mathbb{C}_{1}(0,m)&=&\left\{\begin{array}{ll}
\frac{2}{3} &\mbox{$m=1$}\\
0,&\mbox{$m\neq 1$}
\end{array}
\right.
\end{eqnarray}\normalsize
and
\footnotesize
\begin{eqnarray}
\mathbb{C}_{1}(n,0)&=&\left\{\begin{array}{ll}
\frac{2}{3} &\mbox{$n=1$}\\
0,&\mbox{$n\neq 1$}
\end{array}
\right.
\end{eqnarray}\normalsize
Substituting all the above results in the force expression, we finally have
\footnotesize
\begin{eqnarray}
F_{z}&=&\pi\epsilon\left\{\frac{8}{3}\left(\frac{\epsilon_{p}}{\epsilon_{m}}\right)\frac{k_{0}}{A_{0}C_{0}}R_{1}R_{2}\left(\frac{\sigma}{\epsilon}\right)^{2}\left(3i_{1}B_{01}-\frac{k_{1}}{A_{1}}\mathbb{B}_{10}\right)\right.\nonumber\\
&&-\frac{4}{3}\left(\frac{\epsilon_{p}}{\epsilon_{m}}+2\right)\frac{1}{C_{0}}R_{1}R_{2}\left(\frac{\sigma}{\epsilon}\right)^{2}\left(3i_{1}B_{01}-\frac{k_{1}}{A_{1}}\mathbb{B}_{10}\right)\nonumber\\
&&\left.-\frac{4}{3}\left(\kappa R_{1}\right)^2\frac{k_{0}}{A_{0}C_{0}}R_{1}R_{2}\left(\frac{\sigma}{\epsilon}\right)^{2}\left(3i_{1}B_{01}-\frac{k_{1}}{A_{1}}\mathbb{B}_{10}\right)\right\}\nonumber\\
&=&-4\pi\epsilon\left(\frac{\sigma}{\epsilon}\right)^{2}\left(i_{1}B_{01}-\frac{k_{1}}{3A_{1}}\mathbb{B}_{10}\right)\frac{1}{C_{0}}R_{1}R_{2}\left[-2\left(\frac{\epsilon_{p}}{\epsilon_{m}}\right)\frac{k_{0}}{A_{0}}+\frac{\epsilon_{p}}{\epsilon_{m}}+2+\left(\kappa R_{1}\right)^2\frac{k_{0}}{A_{0}}\right].\nonumber\\
\label{eq:flsapart}
\end{eqnarray}\normalsize
Using expressions (\ref{eq:ABcoeff}) for $\mathbb{B}_{10}$, $A_{0}$ and $A_{1}$, and (\ref{eq:CDcoeff}) for $C_{0}$, we can rewrite the above expression as 
\footnotesize
\begin{eqnarray}
F_{z}&=&4\pi\epsilon\left(\frac{\sigma}{\epsilon}\right)^{2}B_{01}\left\{i_{1}\left(\kappa R_{1}\right)-\frac{k_{1}\left(\kappa R_{1}\right)}{\left(\frac{\epsilon_{p}}{\epsilon_{m}}\right) k_{1}(\kappa R_{1})-\kappa R_{1}k_{1}^{\prime}(\kappa R_{1})}\left[\left(\frac{\epsilon_{p}}{\epsilon_{m}}\right) i_{1}(\kappa R_{1})-\kappa R_{1}i_{1}^{\prime}(\kappa R_{1})\right]\right\}\times\nonumber\\
&&\frac{R_{1}R_{2}}{\kappa R_{2}k^{\prime}_{0}\left(\kappa R_{2}\right)}\left\{2+\frac{\epsilon_{p}}{\epsilon_{m}}-\frac{k_{0}\left(\kappa R_{1}\right)}{\kappa R_{1}k^{\prime}_{0}\left(\kappa R_{1}\right)}\left[\left(\kappa R_{1}\right)^{2}-2\frac{\epsilon_{p}}{\epsilon_{m}}\right]\right\}\nonumber\\
\label{eq:partialLSAforce}
\end{eqnarray}\normalsize
Using Eq.(\ref{eq:Bapprox}), we find
\footnotesize
\begin{eqnarray}
B_{01}(\kappa \mathcal{L})&\approx&\frac{\pi}{2\kappa \mathcal{L}}e^{-\kappa \mathcal{L}}\times\nonumber\\
&&\sum_{\nu=0}^{\infty}A_{01}^{\nu}\left[(\frac{3}{2}-2\nu,0)+(\frac{3}{2}-2\nu,1)\frac{1}{2\kappa \mathcal{L}}\right].\nonumber\\
\label{eq:B01}
&&
\end{eqnarray}\normalsize
According to (\ref{eq:Agamma}),
\footnotesize
\begin{eqnarray}
A_{01}^{\nu}&=&\frac{\Gamma(-\nu+\frac{1}{2})\Gamma(-\nu+\frac{3}{2})\Gamma(\nu+\frac{1}{2})}{\pi\Gamma(-\nu+\frac{5}{2})}\times\nonumber\\
&&\frac{1}{(-\nu)!\nu!}\left(-2\nu+\frac{3}{2}\right).
\label{eq:A01nu}
\nonumber\\
\end{eqnarray}\normalsize
Since the $(-\nu)!$ diverge for all $\nu\in\mathbb{N}$ \cite{abramowitz64}, the only non zero coefficient is
\footnotesize
\begin{eqnarray}
A_{01}^{0}&=&\frac{\Gamma(\frac{1}{2})\Gamma(\frac{3}{2})\Gamma(\frac{1}{2})}{\pi\Gamma(\frac{5}{2})}\frac{1!}{0!1!0!}\left(\frac{3}{2}\right)\nonumber\\
&=&\frac{3}{2}\frac{\Gamma^{2}(\frac{1}{2})\Gamma(\frac{3}{2})}{\pi\Gamma(\frac{5}{2})}=\frac{3}{2}\frac{\Gamma(\frac{3}{2})}{\Gamma(1+\frac{3}{2})}\nonumber\\
&=&\frac{3}{2}\frac{\Gamma(\frac{3}{2})}{\frac{3}{2}\Gamma(\frac{3}{2})}=1
\end{eqnarray}\normalsize
where, in the last line, we have used $\Gamma(z+1)=z\Gamma(z)$. As a result, expression (\ref{eq:B01}) reduces to 
\small
\begin{eqnarray}
B_{01}(\kappa \mathcal{L})&\approx&\frac{\pi}{2\kappa \mathcal{L}}e^{-\kappa \mathcal{L}}\left[(\frac{3}{2},0)+(\frac{3}{2},1)\frac{1}{2\kappa \mathcal{L}}\right].\nonumber\\
&&
\end{eqnarray}\normalsize
According to (\ref{eq:coeffnk}),
\small
\begin{eqnarray}
(\frac{3}{2},0)=(1+\frac{1}{2},0)=\frac{1!}{0!\Gamma(2)}=\frac{1}{\Gamma(1)}=1,\nonumber\\
(\frac{3}{2},1)=(1+\frac{1}{2},1)=\frac{2!}{1!\Gamma(1)}=\frac{2}{\Gamma(1)}=2.\nonumber\\
\end{eqnarray}\normalsize
As a result,
\small
\begin{eqnarray}
B_{01}(\kappa \mathcal{L})&\approx&\frac{\pi}{2\kappa \mathcal{L}}e^{-\kappa \mathcal{L}}\left(1+\frac{1}{\kappa \mathcal{L}}\right).\nonumber\\
&=&\frac{\pi(1+\kappa \mathcal{L})}{2(\kappa \mathcal{L})^{2}}e^{-\kappa \mathcal{L}}.
\label{eq:B01approxf}
\end{eqnarray}\normalsize
Let us now examine the modified Spherical Bessel functions of first kind, which appear in expression (\ref{eq:flsapart}). According to \cite{abramowitz64}, this functions can be written as 
\small
\begin{equation}
i_{n}(x)=(2x)^{-1}\left[R(n+\frac{1}{2},-x)e^{x}-(-1)^{n}R(n+\frac{1}{2},x)e^{-x}\right]
\label{eq:besselfirst}
\end{equation}\normalsize
where
\small
\begin{equation}
R(n+\frac{1}{2},x)=\sum_{k=0}^{n}(n+\frac{1}{2},k)(2x)^{-k},
\end{equation}\normalsize
with $(n+\frac{1}{2},k)$ given by expression Eq.(\ref{eq:coeffnk}). Using this expression along with Eq.(\ref{eq:B01approxf}), $k_{0}(x)=\pi e^{-x}/2x$ and $k_{1}(x)=\pi e^{-x}(1+x)/2x^{2}$ in Eq.(\ref{eq:partialLSAforce}), and performing a lengthy but straightforward calculation, we finally write the force on sphere 1 due to sphere 2 electrical surface charge density and solution electrolytes in the LSA regime as
\small
\begin{eqnarray}
F_{z}&=&-4\pi\epsilon\left(\frac{\sigma}{\epsilon}\right)^{2}\frac{(1+\kappa\mathcal{L})}{\left(1+\kappa R_{1}\right)\left(1+\kappa R_{2}\right)}\frac{(R_{1}R_{2})^{2}}{\mathcal{L}^{2}}\mathcal{F}(\kappa R_{1})e^{-\kappa\left(\mathcal{L}-R_{1}-R_{2}\right)}\nonumber\\
&=&-4\pi\epsilon\psi_{01}\psi_{02}\left(1+\kappa\mathcal{L}\right)\frac{R_{1}R_{2}}{\mathcal{L}^{2}}\mathcal{F}(\kappa R_{1})e^{-\kappa\left(\mathcal{L}-R_{1}-R_{2}\right)},\nonumber\\
&&
\end{eqnarray}\normalsize
\small
where
\begin{eqnarray}
\mathcal{F}\left(\kappa R_{1},\frac{\epsilon_{p}}{\epsilon_{m}}\right)&\equiv&\frac{2+2\kappa R_{1}+\left(\kappa R_{1}\right)^{2}+\left(\kappa R_{1}-1\right)\frac{\epsilon_{p}}{\epsilon_{m}}}{2+2\kappa R_{1}+\left(\kappa R_{1}\right)^{2}+(1+\kappa R_{1})\frac{\epsilon_{p}}{\epsilon_{m}}}\nonumber\\
&&
\end{eqnarray}
and $\psi_{0i},\,i=1,2$, are given by (\ref{eq:psifar}).

\end{chapter}